\documentclass[a4paper, fleqn,usenatbib]{mnras}
\usepackage{multicol}
\usepackage{longtable}
\usepackage{supertabular}
\usepackage{epsfig}
\usepackage{lineno}
\usepackage{lscape}
\usepackage{graphicx}
\usepackage{amssymb}
\usepackage{rotating}
\usepackage{mathtools}
\usepackage{latexsym}
\usepackage{mathrsfs}
\usepackage{natbib}
\usepackage{graphics}
\usepackage{textcomp}
\usepackage{amsmath}
\usepackage{amssymb}

\def\lta{\mathrel{\spose{\lower 3pt\hbox{$\mathchar"218$}} \raise
2.0pt\hbox{$\mathchar"13C$}}} \def\gta{\mathrel{\spose{\lower
3pt\hbox{$\mathchar"218$}} \raise 2.0pt\hbox{$\mathchar"13E$}}}

\title[]{Re-awakening of GRS 1716--249 after 23 years, observed by {\it Swift}/XRT and {\it NuSTAR}}
\author[Bharali et al.]{Priya Bharali$^{1,7}$\thanks{E-mail:priya\_bharali@gauhati.ac.in}, Sunil Chandra$^{2,4}$\thanks{E-mail:sunil.chandra355@gmail.com}, Jaiverdhan Chauhan$^{8,4}$, Javier A. Garc{\'i}a$^{5,6}$, 
\newauthor Jayashree Roy$^{3,9}$, Markus Boettcher$^{2}$ and Kalyanee Boruah$^{1}$ \\
$^{1}$ Department of Physics, Gauhati University, Guwahati-781014, India.\\
$^{2}$ Centre for Space Research, North-West University Potchefstroom, South Africa.\\
$^{3}$ UM-DAE Center for Excellence in Basic Sciences, University of Mumbai, Kalina, Mumbai-400098, India.\\
$^{4}$ Tata Institute of Fundamental Research, Homi Bhabha Road, Mumbai 400005, India.\\
$^{5}$ Cahill Center for Astronomy and Astrophysics, California Institute of Technology, Pasadena, CA 91125, USA.\\
$^{6}$ Remeis Observatory \& ECAP, Universit{\"a}t Erlangen-N{\"u}rnberg, Sternwartstr. 7, 96049 Bamberg, Germany.\\
$^{7}$ Mahatma Gandhi Government Arts College, Mahe, Puducherry-673311, India.\\
$^{8}$ International Centre for Radio Astronomy Research -- Curtin University, GPO Box U1987, Perth, WA 6845, Australia.\\
$^{9}$ Inter-University Center for Astronomy and Astrophysics, Post Bag 4, Pune, Maharashtra, 411007, India.}

\begin{document}

\date{}

\pagerange{\pageref{firstpage}--\pageref{lastpage}} \pubyear{2019}

\maketitle

\label{firstpage}
\begin{abstract}
In this work, we present a spectral and temporal analysis of {\it Swift}/XRT and {\it NuSTAR} observations of GRS 1716--249 during its recent 2016--2017 outburst. This low mass X-ray binary underwent an extraordinary outburst after a long quiescence of 23 years, since its last major outburst in 1993. The source was observed over two different epochs during 2017 April, 07 and 10. The best fit joint spectral fitting in the energy range 0.5 $-$ 79.0 keV indicates that the spectrum is best described by relatively cold, weak disk blackbody emission, dominant thermal Comptonization emission, and a relativistically broadened fluorescent iron K$\alpha$ emission line. We observed a clear indication of a Compton hump around 30 keV. We also detected an excess feature of $\sim1.3$ keV. Assuming a lamp-post geometry of the corona, we constrained the inner disk radius for both observations to 11.92$^{+8.62}_{-11.92}$ R$_{ISCO}$ (i.e., an upper limit) and 10.39$^{+9.51}_{-3.02}$ R$_{ISCO}$ (where R$_{ISCO}\equiv$ radius of the innermost stable circular orbit) for the first epoch (E1) and second epoch (E2), respectively. A significant ($\sim5\sigma$) type$-$C quasi-periodic oscillation (QPO) at $1.20\pm0.04$ Hz is detected for the first time for GRS 1716--249, which drifts to $1.55\pm0.04$ Hz ($\sim6\sigma$) at the end of the second observation. The derived spectral and temporal properties show a positive correlation between the QPO frequency and the photon index. 
  
\end{abstract}


\section{Introduction\label{sec:intro}}

The rapid time variability of transient black hole X-ray binaries (BHXRBs) provides a unique tool to understand the physical processes like accretion, disk geometry, etc., in the vicinity of the central engine. During X-ray outbursts, BHXRBs undergo various spectral state transitions, namely (i) low/hard (LH), (ii) thermal or high soft (HS), (iii) very high state (VHS)/steep power law (SPL), and (iv) hard and soft intermediate (HIMS and SIMS) states \citep{Bel00, Rem06}. The HS state spectrum generally consists of non-thermal flux $\textless$ 25\% of the total emission, while the thermal emission from the accretion disk (around the compact object) constitutes the remaining 75\%, with either a weak or no quasi-periodic oscillations (QPOs) seen in the light curve. On the other hand, the VHS/SPL exhibits peculiar properties such as the disk flux varying from 35\% to 80\%, a photon index ($\Gamma$) $\ge$ 2.4 and a luminosity (L) $\textgreater$ 10\% of the Eddington luminosity (L$_{Ed}$). Low-frequency QPOs are often observed in this state. The LH state is characterized by the disk flux contributing $\textless$ 20\%, and the non-thermal flux dominating at $\ge$ 80\%, a spectral index of 1.4 $\le$ $\Gamma$ $\le$ 2.1, a luminosity L $\leq$ 1\% of L$_{Ed}$, and sometimes low-frequency QPOs ($\textless$ 0.1 Hz) are observed \citep{Rem06}. 

\citet{van89} suggests that the QPO frequencies range from $\sim$mHz to kHz ($\leqslant$1.2 kHz). Based on the peak frequency ($\nu$), QPOs are categorized into low-frequency QPOs (LF--QPOs) and high-frequency QPOs (HF--QPOs). In the case of BHXRBs, the LF--QPOs are characterized by 0.1 $\le$ $\nu$ $\le$ 30 Hz. Using the definition of the Quality factor\footnote{Q--factor is defined as the ratio between the QPO centroid frequency and the full width at half maximum (FWHM) of the QPO peak.}(also referred to as Q--factor), representing the broadness of a QPO peak \citep{Belloni2014}, the LF--QPOs are further sub-categorized into three types: Type--A, Type--B and Type--C \citep{Hom01, Rem02}. Type--A LF--QPOs are characterized by a weak (few percents of rms amplitude) and broad peak (Q--factor $\leq$ 3) around 7--9 Hz, usually observed on top of weak red noise \citep{Hom01, Cas05}. Type--B LF--QPOs represent a relatively strong ( $\sim$4$-$7\% rms amplitude) and narrow peak (Q--factor $\sim$5--7) between 5$-$7 Hz, usually associated with a weak red noise (few percent rms) \citep{Mot11}. The type--C LF--QPOs are characterized by a strong (up to $\sim$21\% rms amplitude) and narrow peak (Q--factor $\sim$5--12) with the variable centroid frequency between 0.1--15 Hz superposed on a strong flat-top noise \citep{Wij99, Motta15}. 

Different models have been proposed to explain the origin and physical nature of QPOs in X-ray binaries (XRBs). The study of LF--QPOs provides an indirect way to understand the accretion flow around the compact object in XRBs. The relativistic precession model (RPM), based on general relativity and proposed by \citet{Ste98}, explains the origin and evolution of LF--QPOs and a few HF--QPOs in neutron star X-ray binaries. According to this model, QPOs originate as a result of the nodal precession, periastron precession, and Keplerian motion, of a luminous blob of material in the accretion flow around the compact object. \citet{Ing09} extended the model proposed by \citet{Ste98}, considering the complete inner flow instead of a luminous blob. The authors tried to demonstrate the origin of LF--QPO and the noise linked to them.  Later \citet{Motta14b} extended this model to black hole X-ray binaries.

The Galactic microquasar GRS 1716--249 (or GRO J1719--24 or Nova Oph 1993) was first detected on 1993 September 25, independently and simultaneously by the BATSE instrument aboard CGRO \citep{Har93} and the SIGMA telescope aboard GRANAT \citep{Bal93}. The optical counterpart was discovered by \citet{del94} and \citet{Mas96}, as a low mass main-sequence star of spectral type K (or later). The mass of the companion is star found to be $\sim 1.6 M_\odot$, and the system has an orbital period of $\sim$ 14.7 hr. The compact object is believed to be a stellar-mass black hole of mass $\sim 4.9 M_\odot$, located at a distance of $\sim$ 2.0--2.8 kpc. The only observed historical major outburst from GRS 1716--249 coincided with its discovery in 1993, which was later studied in detail by \citet{van96}, to investigate the nature of its temporal variability. During the entire 80 days of its outburst, the authors witnessed a QPO at $\nu\sim$ 0.04 Hz at the beginning of the observations, which gradually shifted to 0.3 Hz at the end. A constant phase lag of $0.072\pm0.010$ radian was also observed in the frequency range 0.02$-$0.20 Hz. \citet{van99} observed that the $\le$ 1 Hz QPO was similar to type II burst profiles of Rapid Bursters associated with neutron star accretors. This variability feature suggests that the origin of the $\le$ 1 Hz QPO is independent of the nature of the accretor, and the $\sim$ 0.04 Hz QPO originates from thermal viscous instabilities in the accretion disk surrounding the black hole.

A study of the energy spectra of GRS 1716--249 during the low hard state was presented by \citet{Rev96}, applying different models for the spectral fitting, such as Comptonization and optically thin thermal bremsstrahlung. \citet{Lin05} carried out a detailed study of spectral variability and the soft $\gamma$-ray flux during a 1000 day period in 1993 -- 1995.

Using Monitor of All-sky X-ray Image\footnote{\url{http://maxi.riken.jp/top/index.html}} ({\it MAXI}) data, \citet{Negoro2016} reported the first detection of GRS 1716--249 on 18 December 2016 after $\sim$ 23 years. During the {\it MAXI} observation on 21 December 2016, the source was observed with a photon index of $1.62\pm0.06$ \citep{Masumitsu2016}. {\it Chandra} observations on 06 February 2017 revealed GRS 1716--249 to be in the hard spectral state with a photon index ($\Gamma) \sim 1.53$ \citep{Miller2017}. Based on {\it Swift} observations on 27 March and 2 April 2017, \citet{Armas2017} found that the source was transiting to the soft state. However, {\it Swift} observations on 2017, May 05 and 11 confirmed that the source had returned to the hard state \citep{Bassi2017}. \citet{Bassi2019} studied the 2016--2017 outburst of GRS 1716--249 in the radio and the X-ray bands. Those authors reported that GRS 1716--249 underwent a failed outburst because the source never exhibited the canonical high-soft spectral state during this outburst. The radio -- X-ray correlation shows that the source is positioned at the radio-quiet `outlier' branch \citep{Bassi2019}.

This investigation aims to present a spectro-temporal study of GRS 1716--249 during its recent outburst in 2016--2017, using two different observations performed on 2017 April, 7 and 10 by the {\it Swift}/XRT and {\it NuSTAR} observatories. {\it NuSTAR} offers a pile-up free performance (up to $\sim$100 mCrab), high energy resolution ($\sim$400 eV in the range 0.1$-$10 keV) and excellent calibration. This instrument provides a rare opportunity to study relativistically skewed iron line profiles and to constrain the inner disk radius with high precision \citep{Harrison2013, Miller2013, Pahari2015, Parker2015}. This paper is organized as follows. Section $\S$2 presents a detailed description of the observations and data reduction methodologies. Section $\S$3 describes our analysis and results of the aforesaid outburst. The last section $\S$4 presents our discussion and conclusions of the major obtained results.

\begin{figure*}
\centering
\includegraphics[scale=0.34,angle=-90]{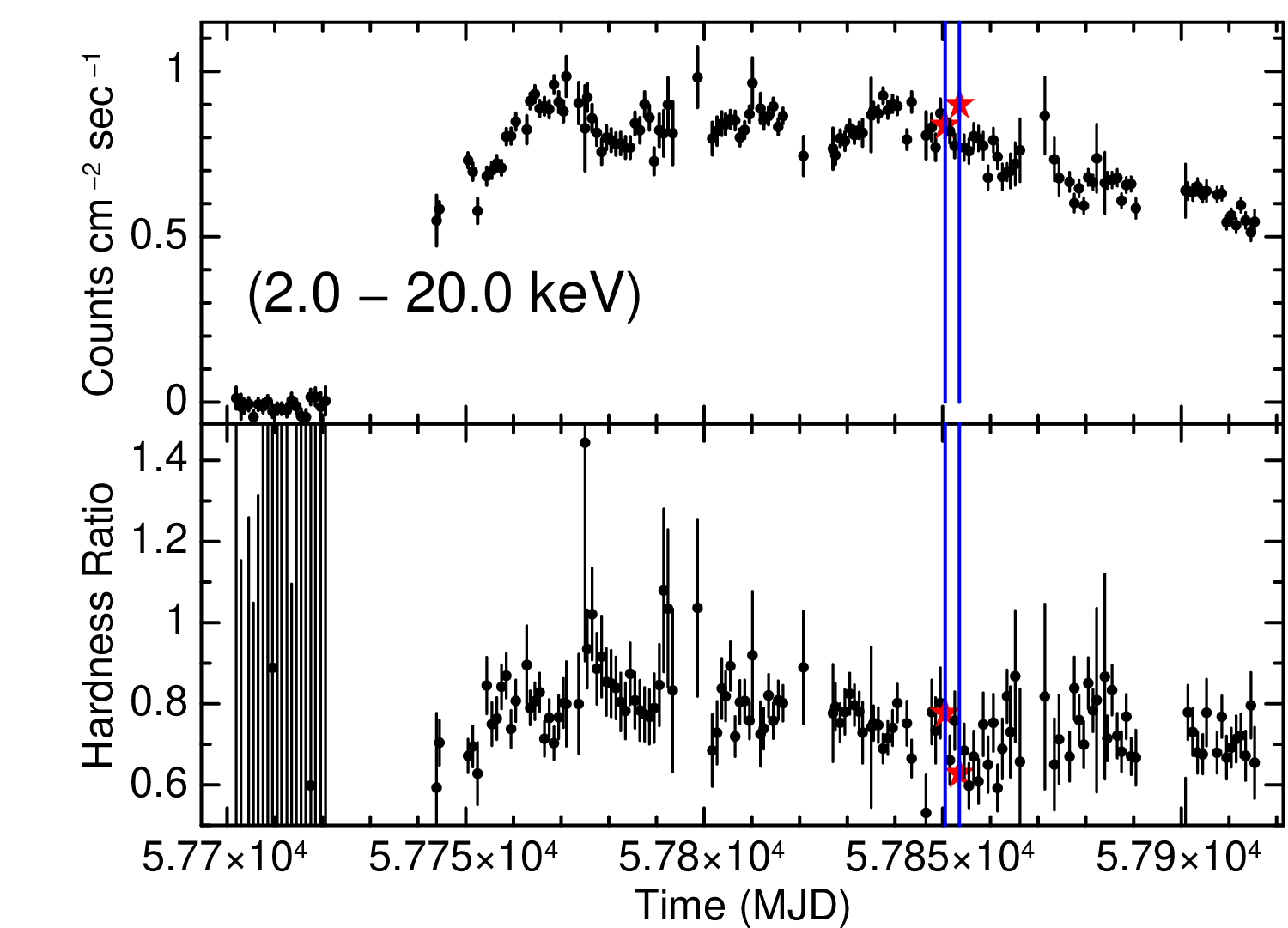}
\includegraphics[scale=0.34,angle=-90]{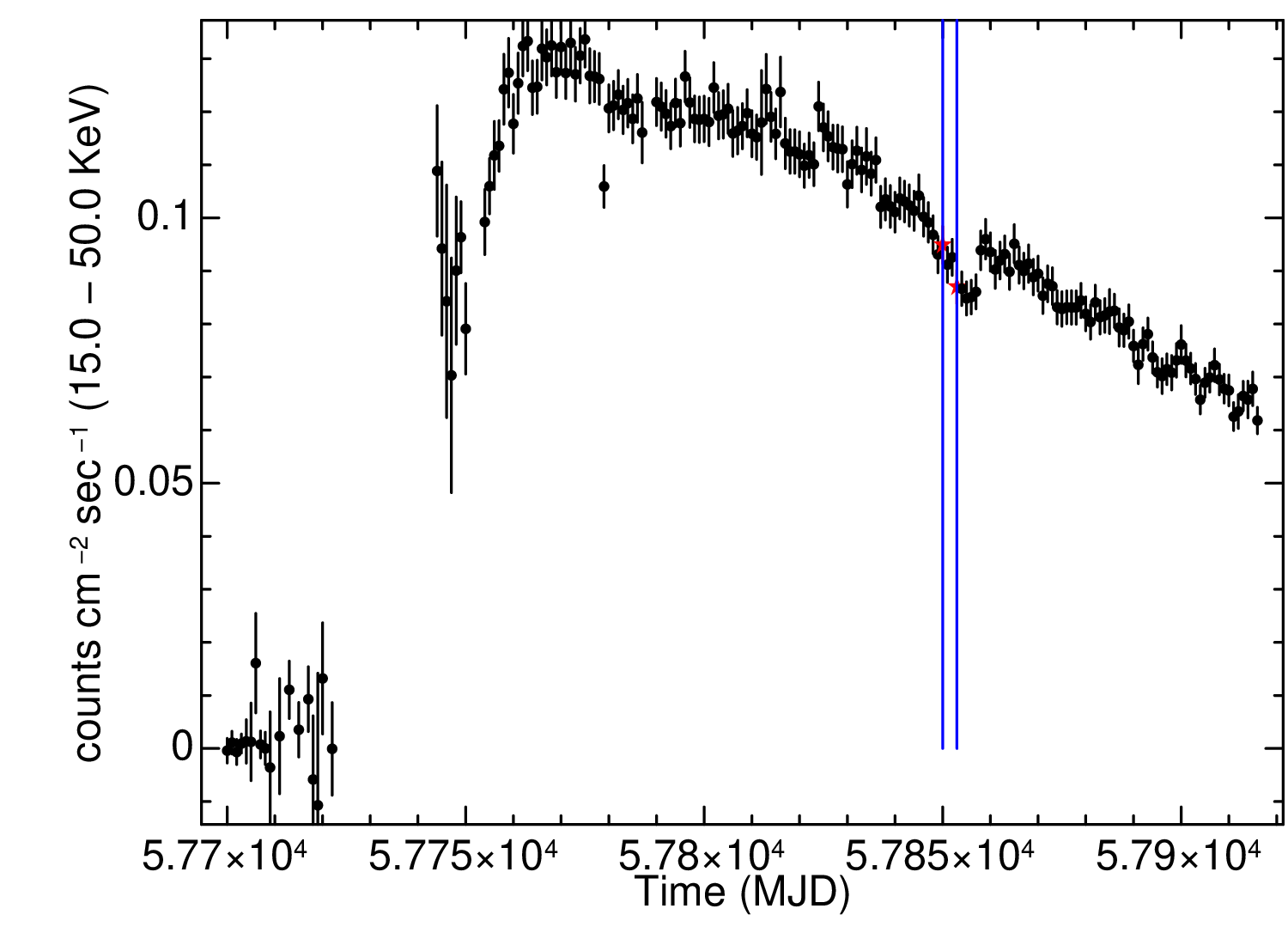}\vspace{2.5em}
\caption{{\bf Left: } One-day averaged light-curve (Top panel) from {\it MAXI} (2.0$-$20.0 keV) along with hardness ratio (Bottom panel) to display the long-term variations in GRS 1716--249. {\bf Right: } The {\it Swift}/BAT light curve in the energy range 15.0-50.0 keV. The blue vertical lines indicate the times of the observations used for the present study. Both observations are highlighted with the red star symbol in the {\it MAXI} and {\it Swift}/BAT plots, and represent the observed flux measured by {\it MAXI} and {\it Swift}/BAT.}
\label{fig:fig1}
\end{figure*}

\begin{figure*}
\centering
\includegraphics[scale=0.33,angle=-90]{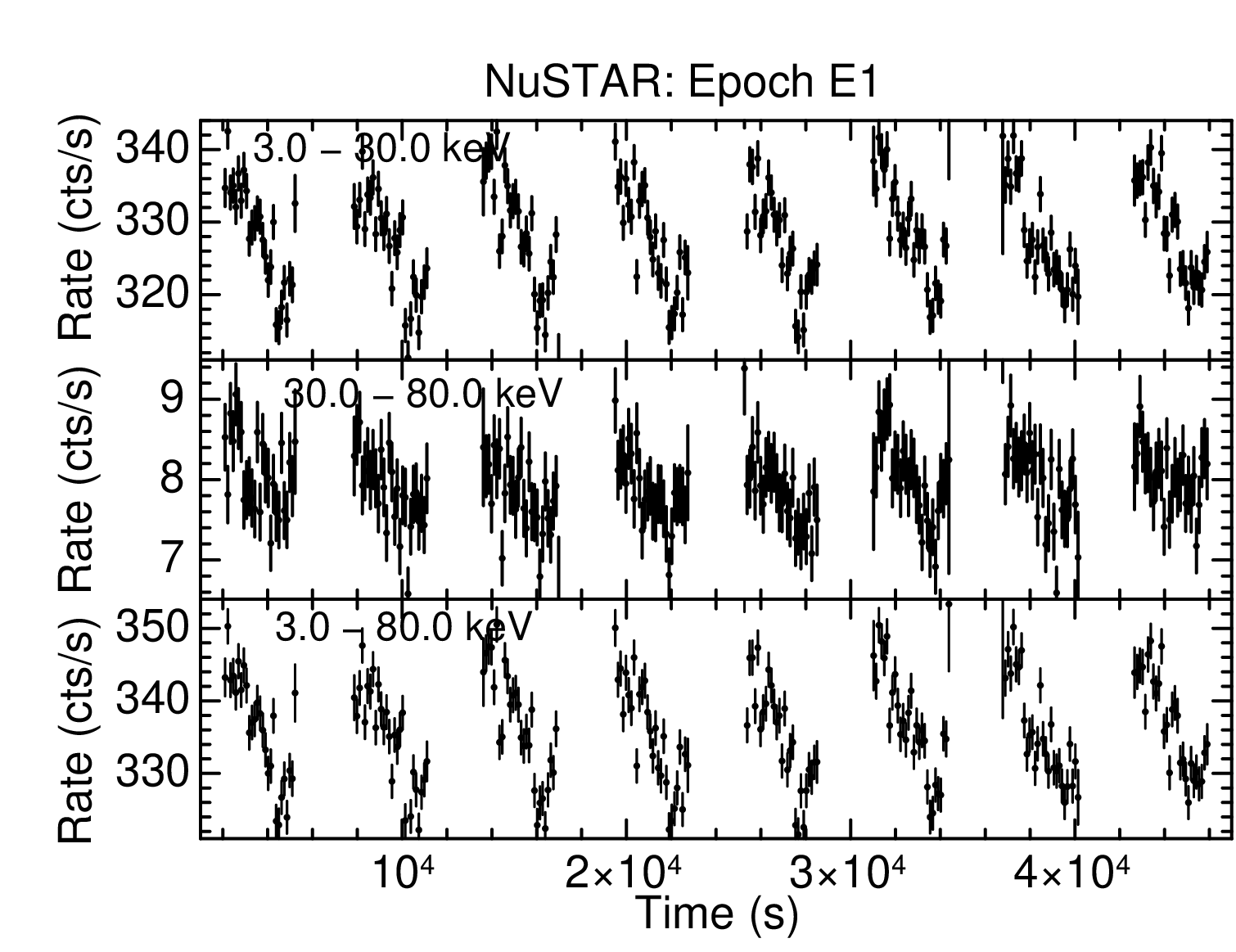}
\includegraphics[scale=0.33,angle=-90]{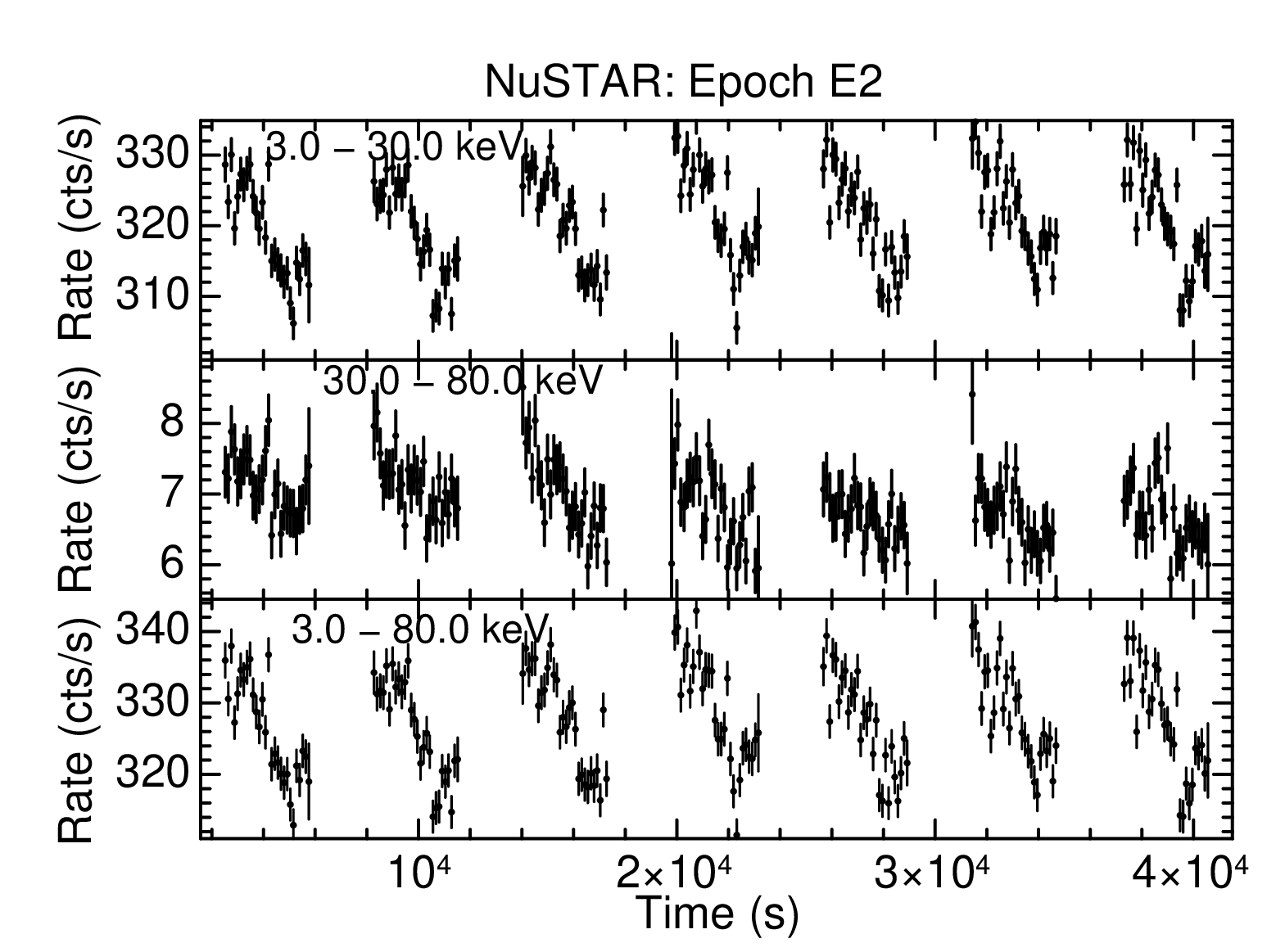}\vspace{1.5em}
\caption{{\bf Left:} {\it NuSTAR} light curves for epoch E1, binned over 120 s, in three different energy bands. The top, middle and bottom panels correspond to the energy bands 3.0-30.0, 30.0-80.0, and 3.0-80.0 keV, respectively. {\bf Right:} Same as the left panel, but for epoch E2 (see Table \ref{tab:tab1} for details of the data). Counts from both FPMA and FPMB are added to improve the $S/N$ ratio.}
\label{fig:fig2}
\end{figure*}


\section{Observation and Data Reduction \label{sec:ObsRedcut}}

\subsection{{\it NuSTAR} FPMA and FPMB \label{subsec:NuSTAR}}

During the 2016 -- 2017 outburst of GRS 1716--249, {\it NuSTAR} observed this source on 2017 April, 07 and 10 (Obs. ID:90202055002 and 90202055004). The {\it NuSTAR} data were acquired using the two focal plane module telescopes (FPMA and FPMB). The net effective on source exposure times after detector dead-time corrections are $\sim$18 ks and $\sim$16 ks, respectively. We refer to the two epochs as E1 and E2 in chronological order. The details of the observations can be found in Table \ref{tab:tab1}. 

\begin{figure*}
\centering
\includegraphics[scale=0.32,angle=-90]{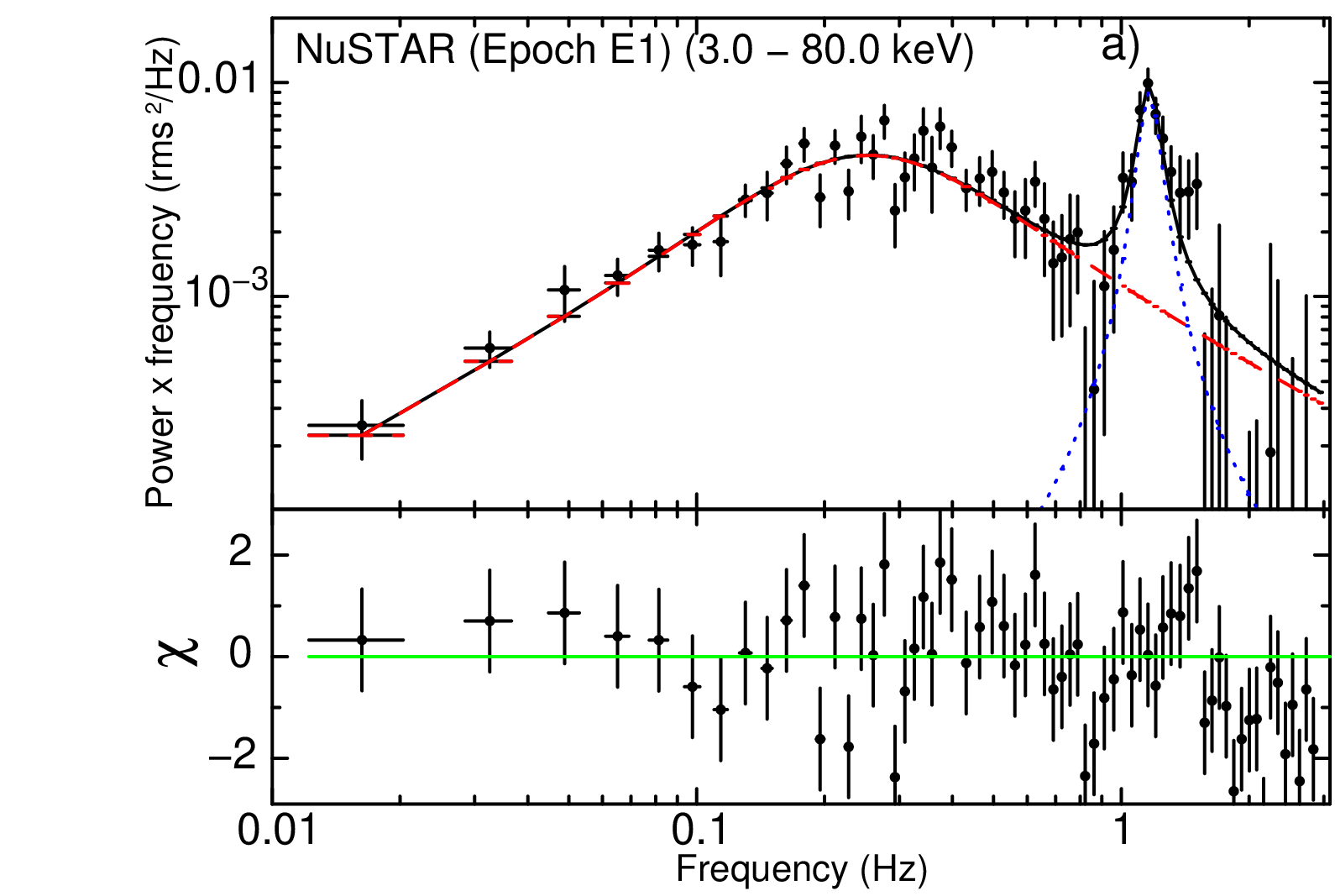}
\includegraphics[scale=0.32,angle=-90]{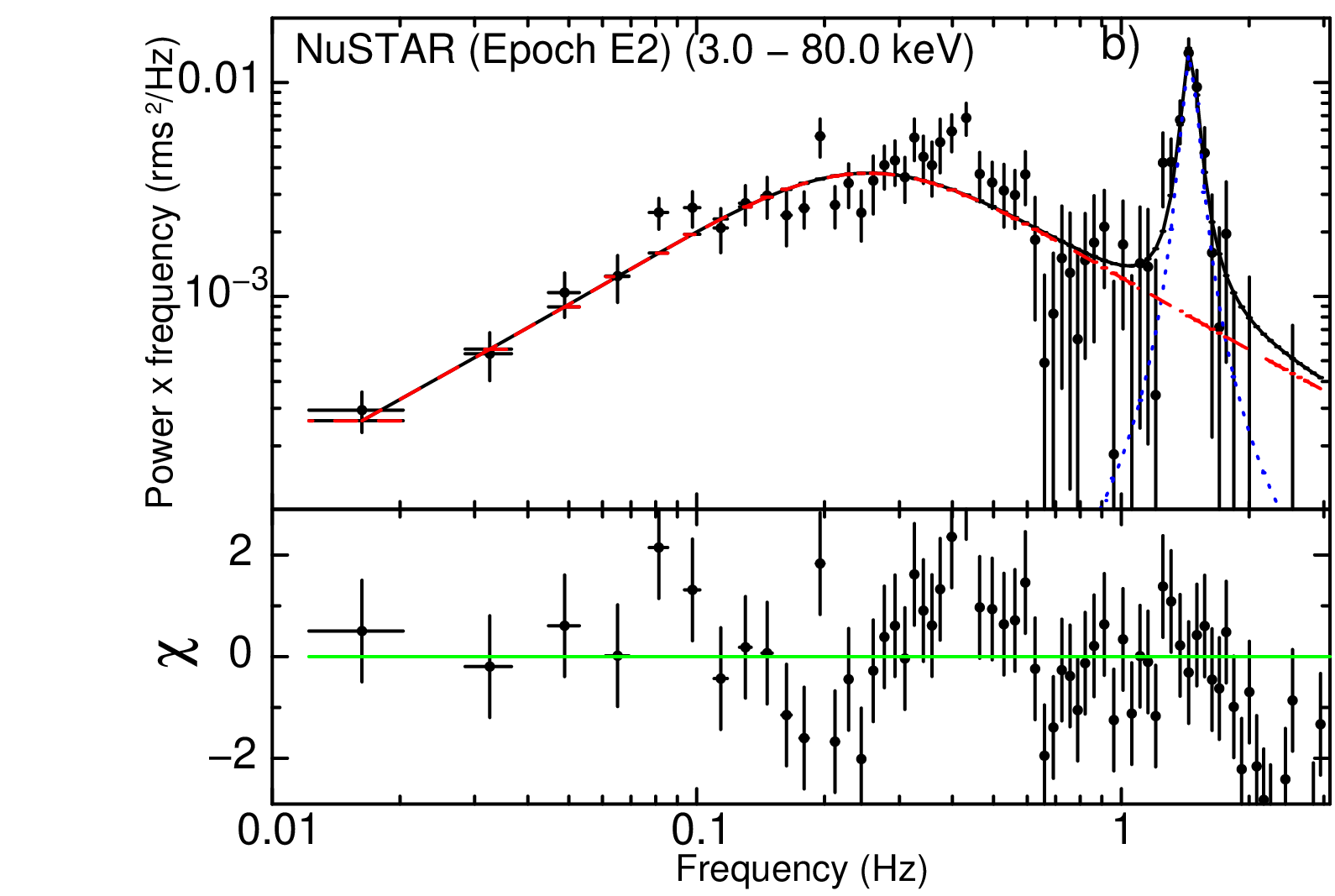}\vspace{3em}
\caption{PDS of the light curves in the 3.0$-$80.0 keV band for epochs E1 (a) and E2 (b), respectively. The bottom panels show the residuals for the same. The red dashed and the blue dotted lines in each plot represent the broad continuum and QPO (peaked at 1.20 Hz for E1 and 1.55 Hz for E2) fitted with a Lorentzian model. The solid black line indicates the total model. The observed QPOs have a significance of 6.9 and 8.1 $\sigma$, respectively.}
\label{fig:fig3}
\end{figure*}

We utilized the standard {\it NuSTAR} data analysis software ({\tt NUSTARDAS} v1.7.1) included in {\tt HEASOFT} v6.23 along with the calibration database {\tt CALDB VERSION 20180419}. We used the {\tt nupipeline} task (version 0.4.6 release date: 2016-10-06) for filtering event files and for making depth corrections from both telescopes. A circular region of radius 100 arcsec centered on GRS 1716--249 was used to extract the source events. For extracting the background events, a circular region of the same size as that of the source with centre 5 arcmin away from the centre of the source was chosen to avoid contamination by the source. Science products, e.g., light curves, energy spectra, response matrix files (rmfs) and auxiliary response files (arfs), for both telescopes (FPMA and FPMB) were generated using the {\tt NUPRODUCTS} task. Light curves from both telescopes were merged to increase the signal-to-noise ratio. To minimize systematic effects, the energy spectra from both detectors were modeled simultaneously. 

\begin{table*}
 \centering
 \caption{Details of our {\it Swift}/XRT and {\it NuSTAR} observations of GRS 1716--249.} 
\begin{center}
\scalebox{0.8}{%
\begin{tabular}{ |l|c|c|c|c|c|c|c|c| }
 \hline
\hline
Instrument  & Observation  & Observation & Observation & MJD & Epoch  & Effective & Count rate & Observation \\
 Name &  ID  & Start date & Start time & & Flag & exposure & (cts s$^{-1}$) & mode \\
  &  & (DD-MM-YYYY) & (hh:mm:ss) & & & time (ks) & &  \\
\hline
{\it NuSTAR}/FPMA, FPMB & 90202055002 & 07-04-2017 & 14:26:09 & 57850 & E1  & $\sim$18 & $338\pm12$ & SCIENCE \\
{\it NuSTAR}/FPMA, FPMB & 90202055004 & 10-04-2017 & 16:36:09 & 57853 & E2  & $\sim$16 & $325\pm10$ & SCIENCE \\
\hline
{\it Swift}/XRT & 00034924029 & 07-04-2017 & 08:50:41 & 57850 & E1  & $\sim$1.6 & $130\pm8$ & WT \\
{\it Swift}/XRT & 00034924031 & 10-04-2017 & 21:12:42 & 57853 & E2  & $\sim$1.9 & $145\pm8$ & WT \\
\hline
\hline
\end{tabular}}\\
\end{center}
\label{tab:tab1}
\end{table*}

\subsection{{\it Swift}/XRT}

The {\it Swift}/X-ray Telescope (XRT) observations on 2017, April 07 and 10 were utilized for the present study. These observations were either exactly simultaneous (E2; Total Exp. = 1.9 ks) or nearly simultaneous (E1; Total Exp. = 1.6 ks; 5 hrs difference from {\it NuSTAR} pointing) with the {\it NuSTAR} observations discussed in \S\ref{subsec:NuSTAR}. During both epochs (E1 and E2), XRT operated in windowed timing (WT) mode. For more details about the observations, please refer to Table \ref{tab:tab1}.

Standard procedures as suggested by the instrument team were followed for the filtering and screening. {\it Swift}/XRT data were reduced using the {\tt xrtpipeline} (version 0.13.2 release date: 2015-01-20). The background-subtracted average count rates during E1 and E2 in WT mode were found to be $\sim$130 and $\sim$145 counts sec$^{-1}$, respectively. According to \citet{Romano2006}, the specified photon pile-up limit of the WT mode data is $\sim$100 counts sec$^{-1}$. Therefore, before extracting any scientific product, the possibility of pile-up in the data is tested and corrected following the prescriptions by \citet{Romano2006}. Specifically, this was done by investigating the spectral distortion due to pile-up and subsequently removing an appropriate bright portion from the center of the source image to restore the spectrum. Following this procedure, the most suitable pile-up free source region used for our analysis is represented by a rectangular region defined by 108$\times$36 arcsec$^2$, with 21.6$\times$21.6 arcsec$^2$ removed from the center. The background region wes chosen to be a similar rectangular region, except 5 arcmin away from the center along the image strip. The end products, namely energy spectrum and light curves corresponding to the source and background regions were extracted using ftools {\tt XSELECT} (V2.4d). Afterwards, the {\tt XRTMKARF} tool was utilized for generating the ARF files using the source spectra and exposure map. The resulting files are then used along with the proper RMF files from the recently updated {\tt CALDB}, for further spectral analysis.   

\section{Analysis and Results}
\subsection{Temporal analysis\label{sec:tempStudy}}

The publicly available one-day averaged {\it MAXI} light curve in the energy range 2.0$-$20.0 keV is plotted in the left panel of Fig. \ref{fig:fig1}, with hardness ratio as a function of time plotted in the bottom. The hardness ratio is defined as the ratio of count rates in the 4.0$-$10.0 keV and 2.0$-$4.0 keV energy bands \citep{Mo13}. The one-day averaged {\it Swift}/BAT light curve in the energy band 15.0$-$50.0 keV is shown in the right panel of the Fig. \ref{fig:fig1}. The long-term light curves from {\it MAXI} (2.0$-$20.0 keV) and BAT (15.0$-$50.0 keV), clearly display the overall rising and declining trends in flux. The average count rate over the two epochs E1 and E2 show a small variation as presented in Table \ref{tab:tab1}. For {\it NuSTAR} (3.0$-$80.0 keV) the average count rate during E1 is slightly higher than during E2, whereas the reverse trend is seen for the average XRT fluxes.

\begin{figure*}
\centering
\includegraphics[scale=0.34,angle=-90]{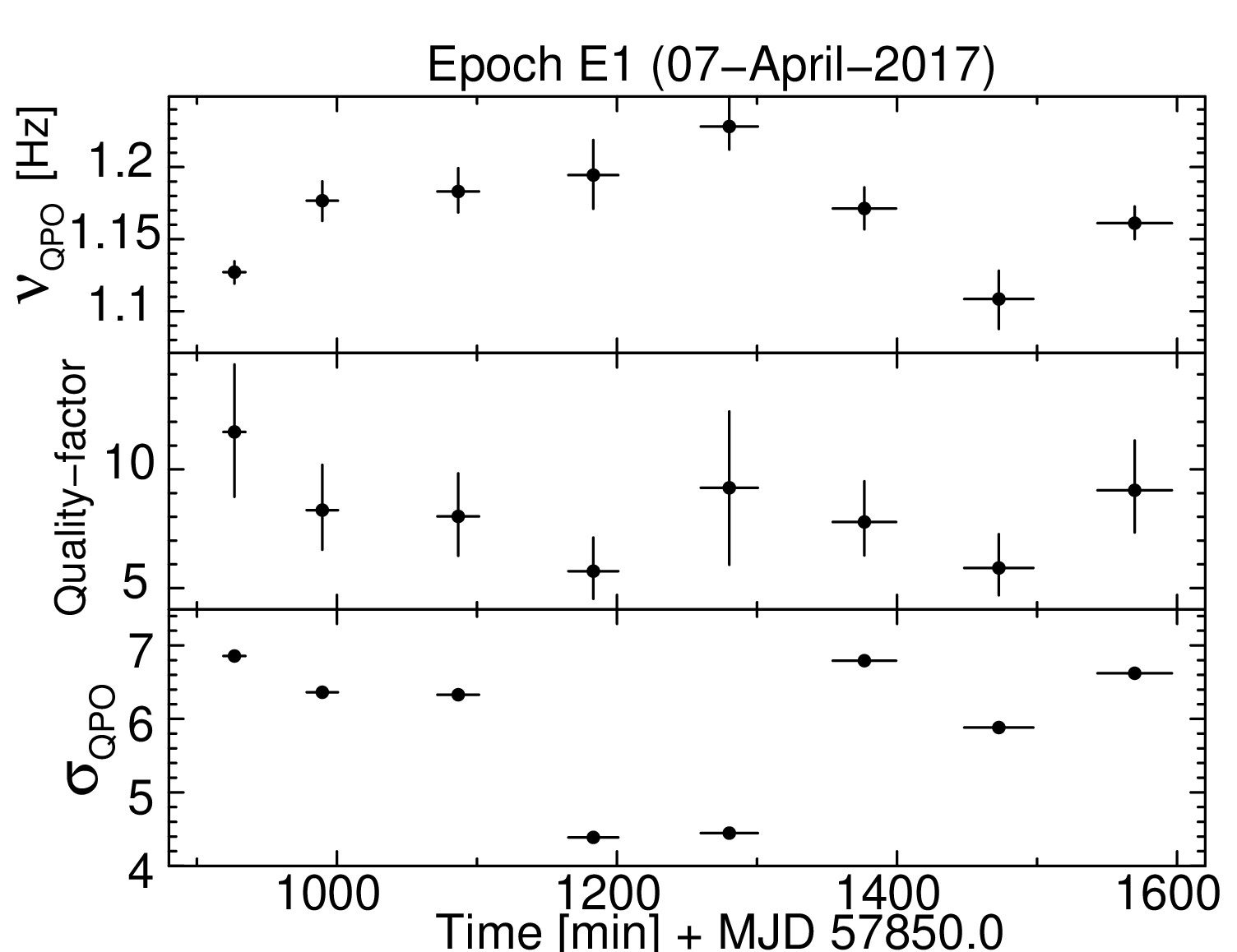}
\includegraphics[scale=0.34,angle=-90]{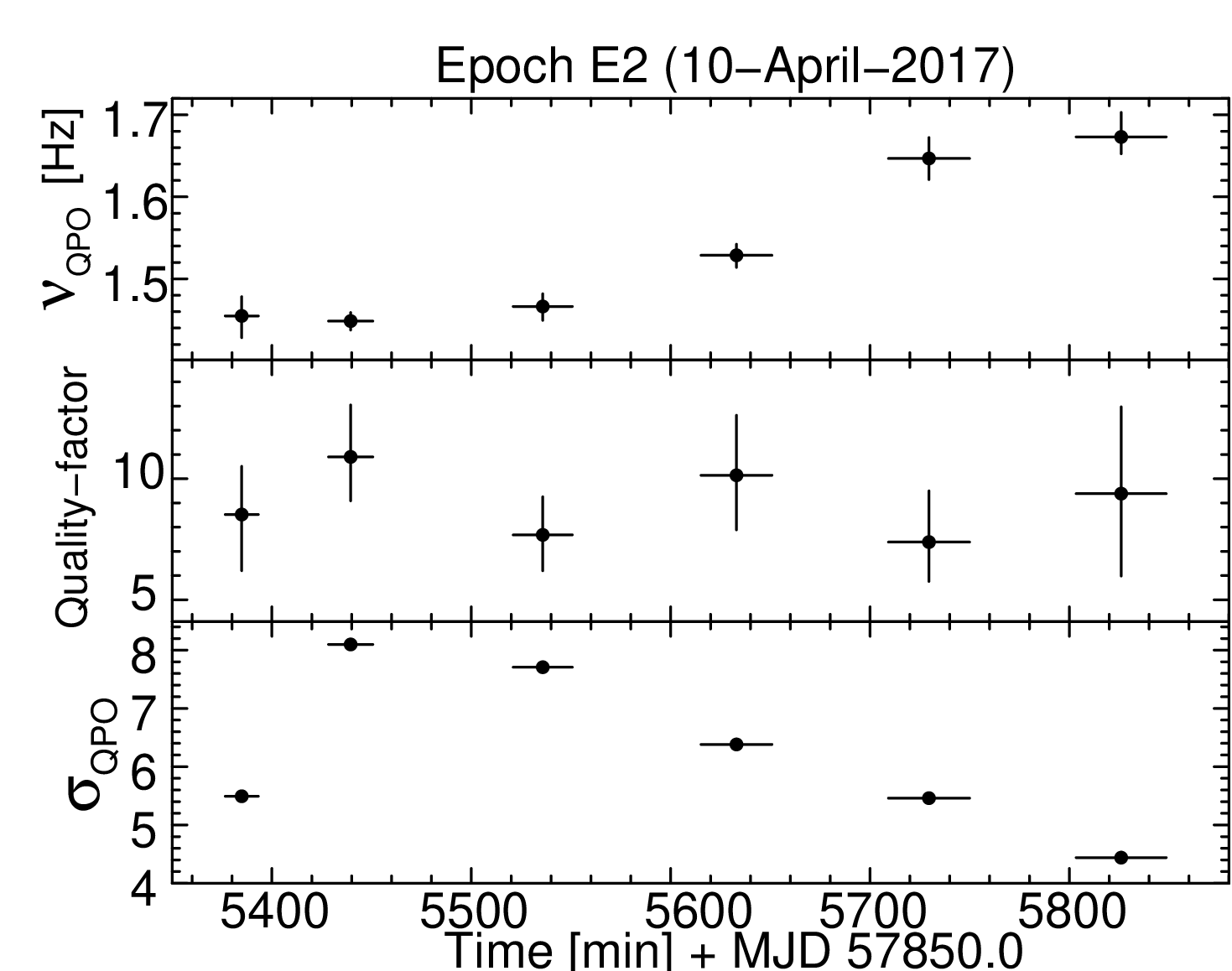}\vspace{1.5em}
\caption{ {\bf Left:} Temporal evolution of the QPO frequency (top panel), Quality factor (middle panel) and QPO significance ($\sigma$; bottom panel) derived from the {\it NuSTAR} observation dated 07 April 2017. {\bf Right :} Same as in the left panel, but for the {\it NuSTAR} observation on 10 April 2017. The QPO frequency clearly shows significant variations in the two epochs, whereas the Q-factor remains constant within uncertainties. The significance of the Q-factor varies, but is always above 4 $\sigma$ (4--7$\sigma$ for E1 \& 4--8$\sigma$ for E2)}
\label{fig:fig4}
\end{figure*}

Fig. \ref{fig:fig2} shows the 120-sec-binned light curve corresponding to E1 and E2 for different energy bands. It indicates periodic variations of $\sim$ 5800 sec, as derived from folding of the light curve over a range of periods and searching for the maximum chi-square as a function of the period in both observations. However, this could be an artifact of the satellite orbital motion. In order to claim a periodicity on this time time-scale, longer observations and rigorous investigations of possible systematics are needed which are beyond the scope for our present work.

To characterize the variability in GRS 1716--249 during both epochs E1 and E2, the power density spectrum (popularly known as PDS) was generated. They are displayed in Fig. \ref{fig:fig3}. PDS are generated by using the Fourier transform of the light curves. The peaks in the PDS correspond to the presence of periodic signals in the light curves. The {\tt powspec} tool distributed as {\tt XRONOS} sub-package of the heasoft package is used to generate the PDS. The {\it NuSTAR} detectors are affected by dead-time, which is $\approx$ 2.5 ms \citep{Harrison2013, Bachetti2015}. However, we generated PDS for FPMA and FPMB separately, which are not dead-time corrected. The presence of dead-time affects the contribution of white noise in the PDS and sometimes affects the overall shape of the QPO, but there is no effect on the QPO peak frequency. A distortion of the overall QPO shape affects the Q-factor and significance, but the variation is only significant at higher count rates, above $\sim$600 counts/sec \citep{Bachetti2015}. We observed a QPO in both detectors. The PDSs were fitted with a model consisting of two Lorentzians added together. The first, broader Lorentzian component represents the continuum, whereas the second Lorentzian was used to fit the QPO. The significance of the QPOs was estimated by the ratio of the area under the Lorentzian profile peaking at the QPO frequency (shown with the blue dash-dot-dashed line in Fig. \ref{fig:fig3}) to the 1$\sigma$ negative error in the area estimated by the model. The recipe suggested by \citet{Vau05} was adopted to calculate the confidence level of the detected QPO and to reject low significance peaks.

In Table \ref{tab:tab3}, we present the derived broad Lorentzian knee frequency, QPO frequency, Q-factor, its significance and $\chi^{2}$/dof for various segments of the {\it NuSTAR} light curves for both epochs in columns 3, 4, 5, 6 and 7, respectively. The temporal evolution of these parameters is shown in Fig. \ref{fig:fig4}. It is evident from Fig. \ref{fig:fig4} that there is a hint of variations in the QPO peak frequency ($\nu_{QPO}$) during both epochs, and we measured a significant change of the QPO peak frequency between the two epochs. During E1, it rises consistently from 1.13 Hz to 1.23 Hz in $\approx$ 6 hrs and afterward decreases to $\approx$ 1.11 Hz within the next $\approx$ 3 hrs, before rising again as evident from the last segment. Although the Q-factor shows some signature of variations, it always remains within 1$\sigma$ uncertainty. The significance of the QPO detection ($\sigma_{QPO}$) was always above 4. The beginning of E2 witnessed a higher value of $\nu_{QPO}$ of 1.45 Hz which remained constant within 1$\sigma$ for approximately 2.5 hours. The next four segments indicate a rising trend for the rest of the observations. Q-factor seems to be constant throughout E2. We also find that the QPO detection is always significant at $\ge 4 \, \sigma$. Note that the two epochs E1 and E2 are separated by approximately 2.6 days. Therefore, there is an indication that on an average $\nu_{QPO}$ is drifting towards higher frequencies with temporary reversals of the trend inbetween. However, the Q-factor shows a small hint of variations with a mean value $\approx$ 8 within 2$\sigma$ errors.

Type--C QPOs are strong (rms amplitude $\sim$ 21 \%), narrow (Q--factor $\sim$ 4--12) with centroid frequency varying within 0.1$-$15 Hz and superimposed on a strong flat-top noise \citep{Cas05, Pahari2014}. As displayed in Fig. \ref{fig:fig4} and evident from the PDS fitting parameters listed in Table \ref{tab:tab3}, the QPO frequency varies in the range 1.11 -- 1.67 Hz, the Q-factor ranges within $\sim$ 5--12 and a broad Lorentzian knee frequency is also present in our observations. This confirms that the observed low-frequency QPO to be of type--C. The PDS plots displayed in Fig. \ref{fig:fig3} also confirm the same type of low-frequency QPO. Both the PDS are Leahy-normalized and noise subtracted.

\begin{figure*}
\centering
\includegraphics[scale=0.33,angle=-90]{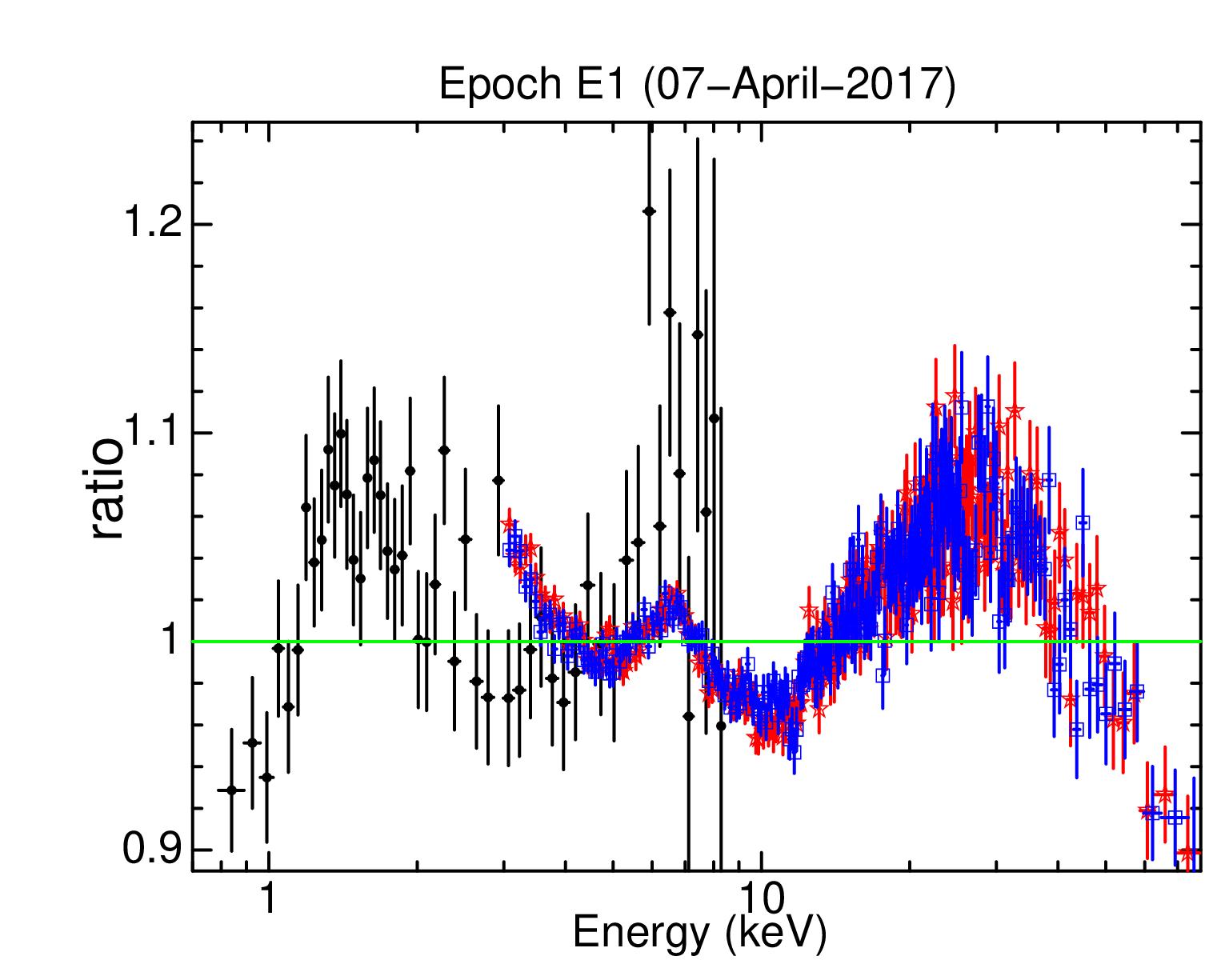}
\includegraphics[scale=0.33,angle=-90]{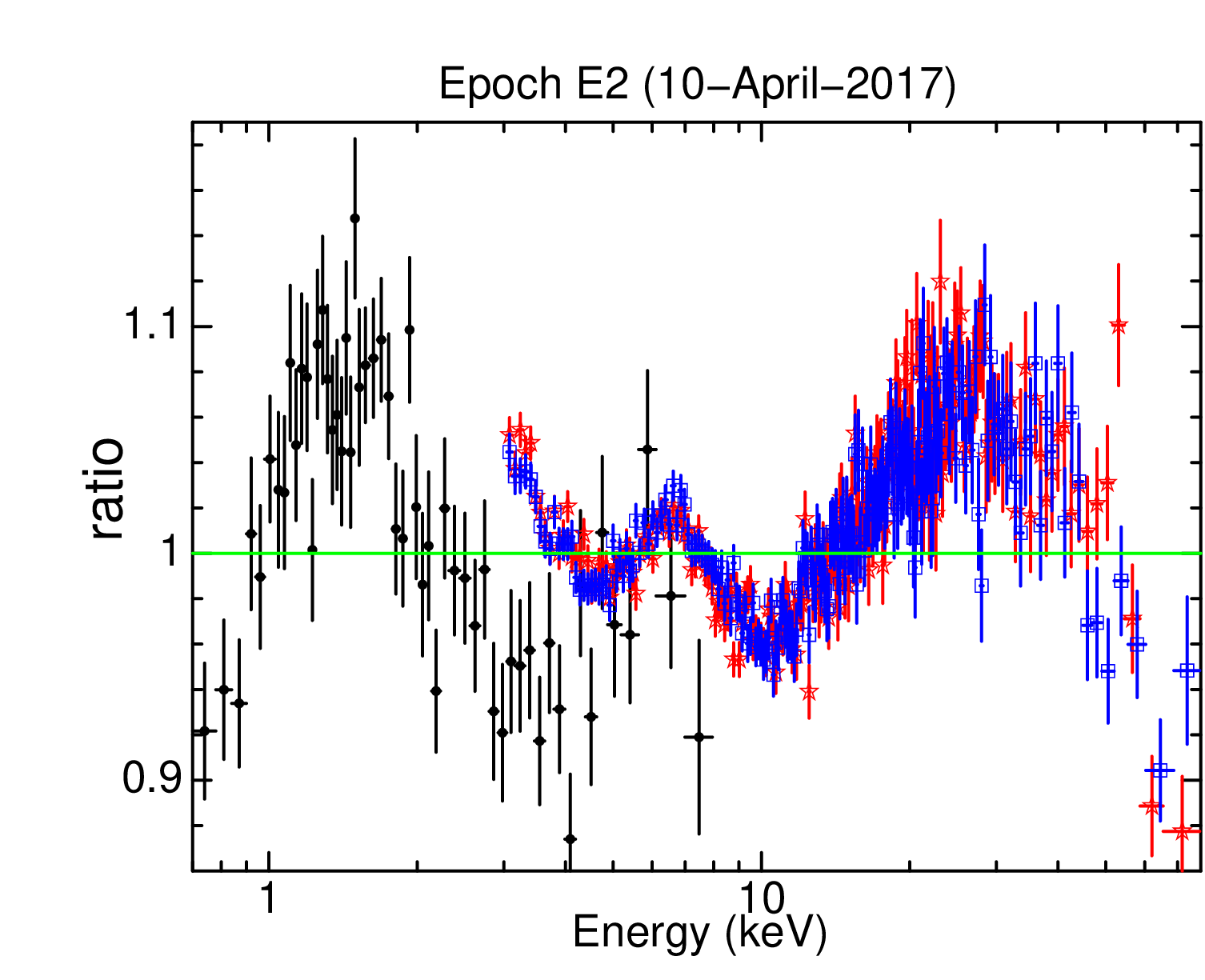}\vspace{1.5em}
\caption{{\bf Left:} Ratio of jointly fitted Swift/XRT (black dots) and {\it NuSTAR} FPMA (red star)/FPMB (blue square) spectra for epoch E1. The spectra are fitted using a simple phenomenological model {\tt TBabs$\times$(diskbb+nthComp)} for observation E1 in the energy range 0.5$-$79 keV. A broad emission line feature in the energy range 5$-$8 keV and a dip around 11 keV is significantly detected. Around 1 and 30 keV a hump-like excess is also observed. {\bf Right:} Same as on the left, but for epoch E2.}
\label{fig:fig5}
\end{figure*}

\subsection{Spectral analysis\label{sec:specStudy}}

To understand the spectral characteristics of GRS 1716--249 during the epochs E1 and E2, a simultaneous fit of the broad-band spectrum from {\it NuSTAR}/FPMA, FPMB (3.0$-$78.0 keV) and {\it Swift}/XRT (0.5$-$8.0 keV) was performed using {\tt XSPEC} version: 12.10.0 \citep{Arn96}. Before employing more complex models, the spectral fitting was first attempted with individual standard continuum models like {\tt diskbb}, {\tt nthComp} and {\tt powerlaw}. They generally result in large reduced chi-square values ($\chi^2_\nu$) ($\gg$ 2; i.e., greater than the acceptable limit). 


An additional constant multiplicative term was incorporated in the models, utilizing the {\tt CONSTANT} model inbuilt in {\tt XSPEC}, to adjust the factors related to the cross-instrument calibration uncertainties. This constant is kept fixed at 1 for {\it Swift}/XRT, and the spectra were fitted by keeping it free for FPMA and FPMB. In this way, the best-fit constants for FPMA and FPMB represent the relative cross-instrument calibration factors with respect to {\it Swift}/XRT. This relative cross-instrument factor for E1 was estimated as $1.16\pm0.02$ ($\sim$16\%) for both FPMA and FPMB, while for E2 it was found to be $1.08\pm0.01$ ($\sim$8\%) for both telescopes. The recommended acceptable range for the cross-instrument calibration factor between {\it Swift} and {\it NuSTAR} is 3$-$20 \% \citep{Mad15, Mar17}. Hence, our values clearly fall within the acceptable range as recommended in the literature.


\begin{figure*}
\centering
\includegraphics[scale=0.32,angle=-90]{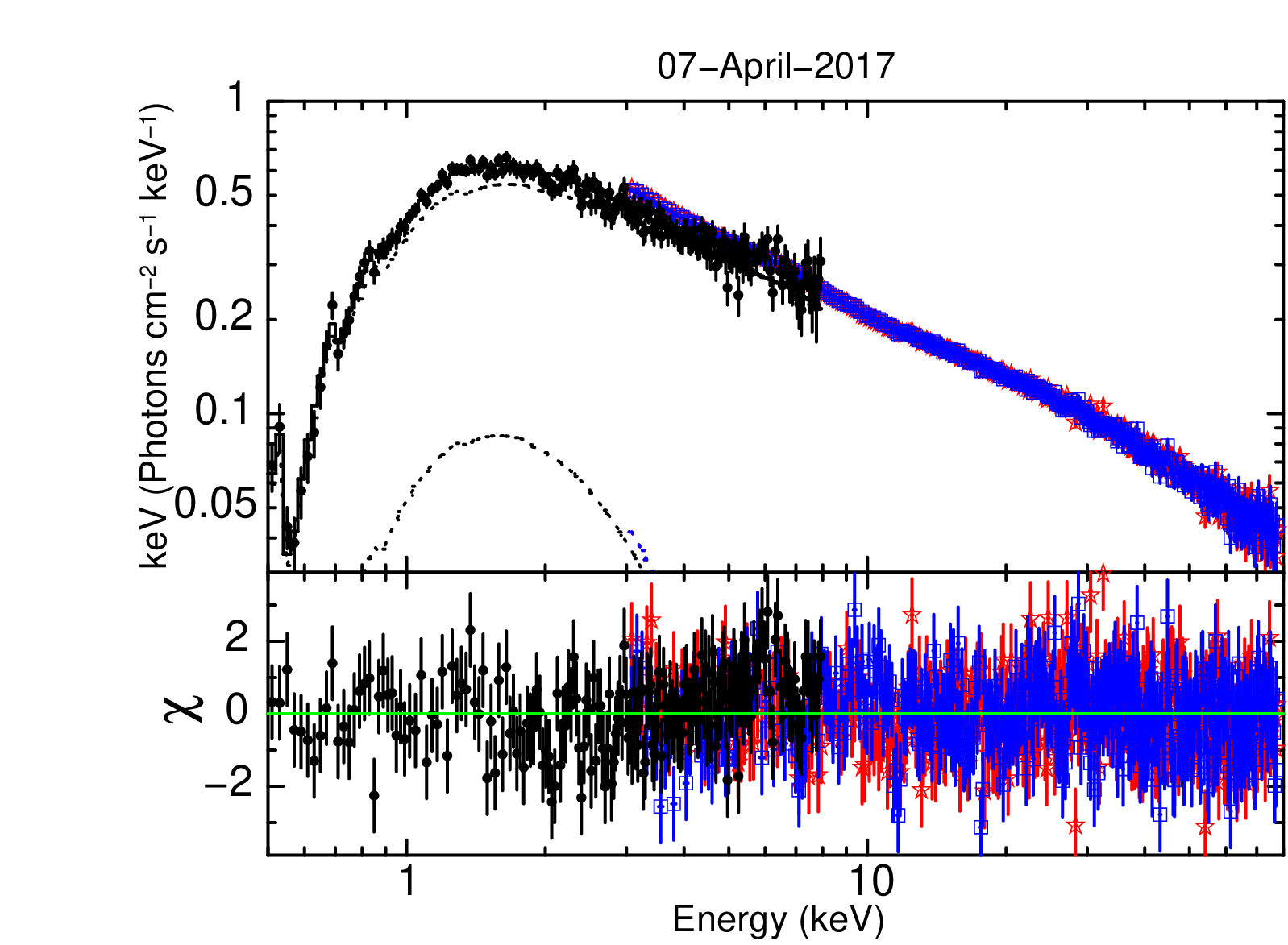}
\includegraphics[scale=0.32,angle=-90]{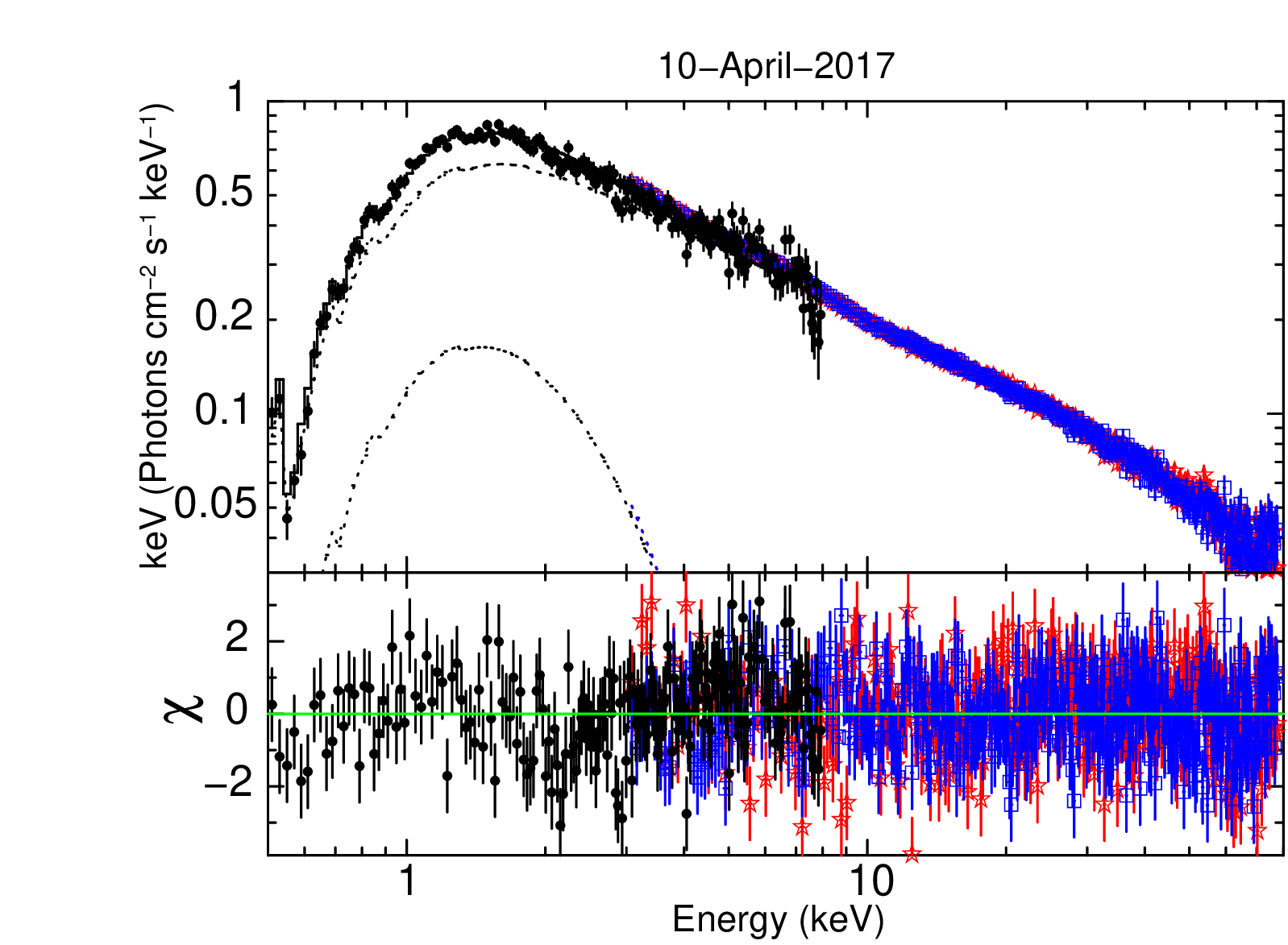}\vspace{2em}
\caption{Swift/XRT (black dots) and {\it NuSTAR} FPMA (red star)/FPMB (blue square) unfolded spectra of GRS 1716--249 along with the best fit model components {\tt (TBabs$\times$(diskbb+relxill(lp)Cp))}, and residuals for epoch E1 ({\bf Left panel}) and epoch E2 ({\bf Right panel}). a$^{*}$ is fixed at 0.998.}
\label{fig:fig6}
\end{figure*}

To consistently explain the observed broad-band spectrum, two-component models such as {\tt TBabs$\times$(diskbb+nthComp)}, {\tt TBabs$\times$(diskbb+powerlaw)} and {\tt TBabs$\times$(nthComp+powerlaw)} were further fitted to the data. The observed spectrum at both epochs (E1 \& E2) were found to be reasonably well described by the {\tt TBabs$\times$(diskbb+nthComp)} model, i.e, the combination of multi-colored disk blackbody component \citep[{\tt diskbb:}][]{Mitsuda1984, Makishima1986} and a thermal Comptonization model \citep[{\tt nthComp:}][]{Zd96,Zy99} modulated by Galactic absorption by neutral hydrogen \citep[{\tt TBabs:}][]{Wil00}. We found excesses around 1.3, 6.4 and 30 keV as shown in Fig. \ref{fig:fig5}. An excess around 6.4 keV is indicative of an Fe emission line from the accretion disk, while the observed residuals around 30 keV indicate a Compton hump, which, together with the relativistically broadened Fe K$\alpha$ line, is typical for Compton reflection. A soft excess around 1.3 keV is typically observed when the surface of the accretion disk is ionized. In this case, electron scattering becomes important along with the re-emission of emission lines, causing a soft excess \citep{Ross1993, Ross2005, Garcia2013}. We found that the red wing of the Fe emission line is extended down to 6 keV while the blue wing is stretched close to 8 keV with a dip around 11 keV as shown in Fig. \ref{fig:fig5}.

Therefore, we replaced the {\tt nthComp} model by the self-consistent, broadband reflection model {\tt relxilllpCp} from the {\tt relxill} \citep[v1.0.2 :][]{Garcia2014, Dauser2014} model family, to account for the relativistic reflection spectrum. Thus, our best fit model is {\tt TBabs$\times$(diskbb+relxilllpCp)}. The {\tt relxilllpCp} model is based on a lamp-post geometry of the corona, which is believed to be the illuminating source. In the lamp-post geometry, the corona is treated as a point source positioned at a height h, on the black hole spin axis above the accretion disk. The {\tt relxilllpCp} model uses the thermal Comptonization model {\tt nthComp} as the input continuum. In the case of the {\tt relxilllpCp} model, the value of the reflection fraction (Refl$_{frac}$) can be self-consistently determined, depending on the values of the lamp-post height (h), the black hole spin parameter (a$^{*}$) and the inner accretion disk radius (R$_{in}$) through ray-tracing calculations. The evaluation of Refl$_{frac}$ by the model itself helps to reduce the parameter space and eventually constrains the geometry of the system \citep{Dauser2014}.

While fitting the average spectra for both epochs using best fit model (i.e., {\tt TBabs$\times$(diskbb+relxilllpCp)}), we fixed the value for the outer radius of the accretion disk at 400 r$_{g}$, where r$_{g}$ is the gravitational radius (r$_{g} \equiv$ GM/c$^{2}$). We simultaneously fitted for a$^{*}$ and R$_{in}$ but those two parameters are degenerate, and the effective inner accretion disk radius is controlled by both a$^{*}$ and the disk inclination angle. In the case of a non-rotating black hole (a$^{*}$ $\approx$ 0), R$_{ISCO} \equiv$ 6r$_{g}$. However, in the case of a Kerr black hole, considering the co-rotating case where the accretion disk is rotating in a direction same as the compact object, R$_{ISCO} \equiv$ r$_{g}$ \citep{Bardeen1972, Thorne1974}. During the spectral fitting for both E1 and E2, the spin parameter (a$^{*}$) approached the hard upper limit of 0.998. Therefore, we have frozen the value of a$^{*}$ at 0.998 and kept R$_{in}$ free to vary. For a black hole of a$^{*}$ $\approx$ 0.998, R$_{ISCO}$ $\approx$ 1.24 r$_{g}$ \citep{Bardeen1972, Thorne1974}.

Additionally, to determine the real nature of the system i.e., whether the compact object is favouring a low spin or a truncated disk, we fixed the value of R$_{in}$ at the ISCO and allowed the spin parameter to vary. We found that a$^{*}$ continued to saturate at the maximum value and the 2$\sigma$ uncertainty on the spin parameter provided a lower bound of 0.73. It is observed that the $\chi^{2}$/dof is 949.04/906 if we freeze a$^{*}$ at its maximum value (0.998), whereas it is 1010.72/906 when we fix R$_{in}$ at the ISCO. This implies that the system prefers a truncated disk over a low spin.

During our analysis for both epochs, we were unable to constrain the electron temperature (kT$_{e}$). Therefore, we fixed it at a value of 400 keV. From our spectral fittings, the iron abundance (A$_{Fe}$) was found to be 0.79$^{+0.11}_{-0.05}$ times the solar abundance for E1. However, for E2 the value of the iron abundance is found to be 1.01$^{+0.34}_{-0.08}$ times the solar abundance. Apart from the above-mentioned parameters, we determined the following reflection and continuum parameters using the {\tt relxilllpCp} model: the lamp-post height (h), the power-law index ($\Gamma$), the R$_{in}$, the ionization index of the accretion disk (log $\xi$) and the flux due to reflection.

Even after using the best fit model ({\tt TBabs$\times$(diskbb+relxilllpCp)}), we have observed a small excess around 6-7 keV in the XRT spectra. The main reason for this excess could be cross-calibration errors. The position of the line is better constrained by \textit{NuSTAR} spectrum in comparison to the \textit{Swift}/XRT spectrum, due to the high signal-to-noise ratio for \textit{NuSTAR}, owing to its excellent sensitivity in this energy range. The iron line profiles observed by \textit{Swift}/XRT and \textit{NuSTAR} do not match perfectly for the same reason, and hence identical constraints for the Gaussian profile of the line for both instruments cannot be obtained by simultaneous fitting. Therefore, a small excess around 6--7 keV is visible in the \textit{Swift}/XRT spectrum. The equivalent widths of the 6.4 keV line derived from simultaneous fitting, corresponding to the epochs (E1) and (E2), are found to be $0.222\pm0.001$ keV and $0.173\pm0.001$ keV, respectively.

The fitted spectra along with the different model components and residuals, for both epochs (E1 \& E2) are presented in the left and right panel of Fig. \ref{fig:fig6} (See. Table \ref{tab:tab2} for the best-fit parameters), respectively. For the broad-band continuum, we used the {\tt TBabs} \citep{Wil00} model inbuilt in {\tt XSPEC} to account for the Galactic neutral hydrogen column density. The most recent abundance model \citep[{\it aspl}:][]{Aspl2009}, inbuilt in {\tt XSPEC}, was adopted as an abundance input for the {\tt TBabs} model. The best-fit absorption column density, $n_H$ (in units of 10$^{22}$ cm$^{-2}$) was found to be 0.62$^{+0.02}_{-0.02}$ and 0.59$^{+0.01}_{-0.01}$ for the epochs E1 and E2, respectively. These values show significant excess over the Galactic value in the direction of source as estimated by the online tools of the LAB Survey\footnote{\url{https://www.astro.uni-bonn.de/hisurvey/profile/index.php}} \citep{Kalberla2005} (n$_{H} = 0.26 \times 10^{22} cm^{-2}$) implying the presence of intrinsic excess absorption in this BHXRB system. 

The disk temperature (T$_{in}$ of the {\tt diskbb} model) shows a small variation but the ionization index of the accretion disk (log $\xi$), as well as the lamp-post height (h) as estimated from the broad-band spectral fitting at E1 and E2, do not show any significant change. The power-law index ($\Gamma$) increases from 1.768$^{+0.013}_{-0.006}$ (E1) to 1.812$^{+0.006}_{-0.006}$ (E2). This increment in $\Gamma$ suggests that the source is drifting towards the soft state. The obtained inner disk radii R$_{in}$ (r$_{g}$) for E1 and E2 are 14.78$^{+5.93}_{-14.78}$ r$_{g}$ and 12.88$^{+5.06}_{-3.58}$ r$_{g}$, respectively. The disk ionization (log $\xi$) is high for both observations and its values corresponding to E1 and E2 are found to be 3.07$^{+0.06}_{-0.03}$ and 3.19$^{+0.28}_{-0.07}$, respectively. The best-fit lamp-post height (h) is 23.77$^{+16.02}_{-10.84}$ and 12.75$^{+7.30}_{-5.11}$ in units of r$_{g}$. The variation in h is small and within 2$\sigma$ error bars. The 2$\sigma$ error in all the above-mentioned quantities were calculated using the error command in {\tt XSPEC}.

\begin{figure*}
\centering
\includegraphics[scale=0.33,angle=-90]{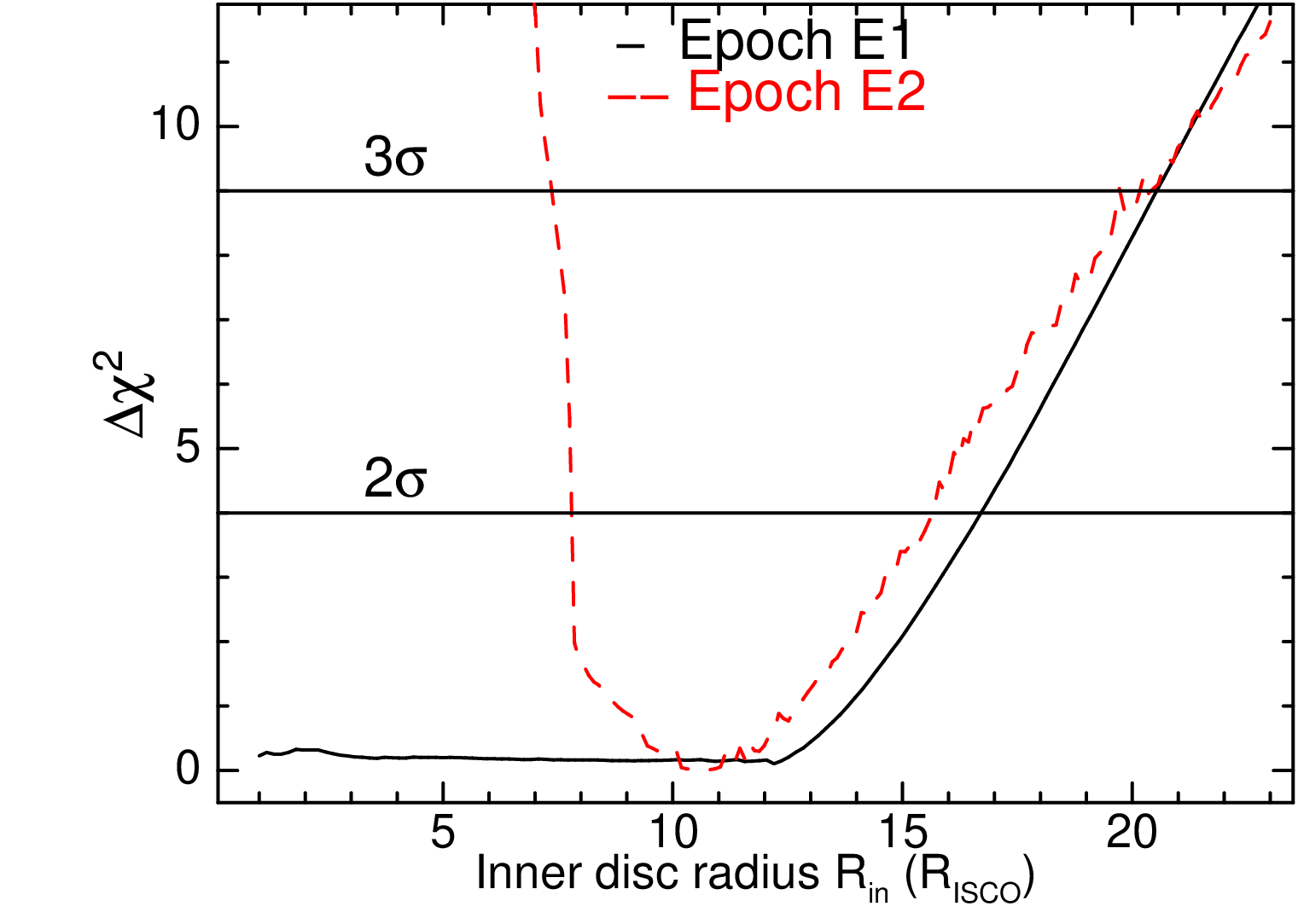}
\includegraphics[scale=0.33,angle=-90]{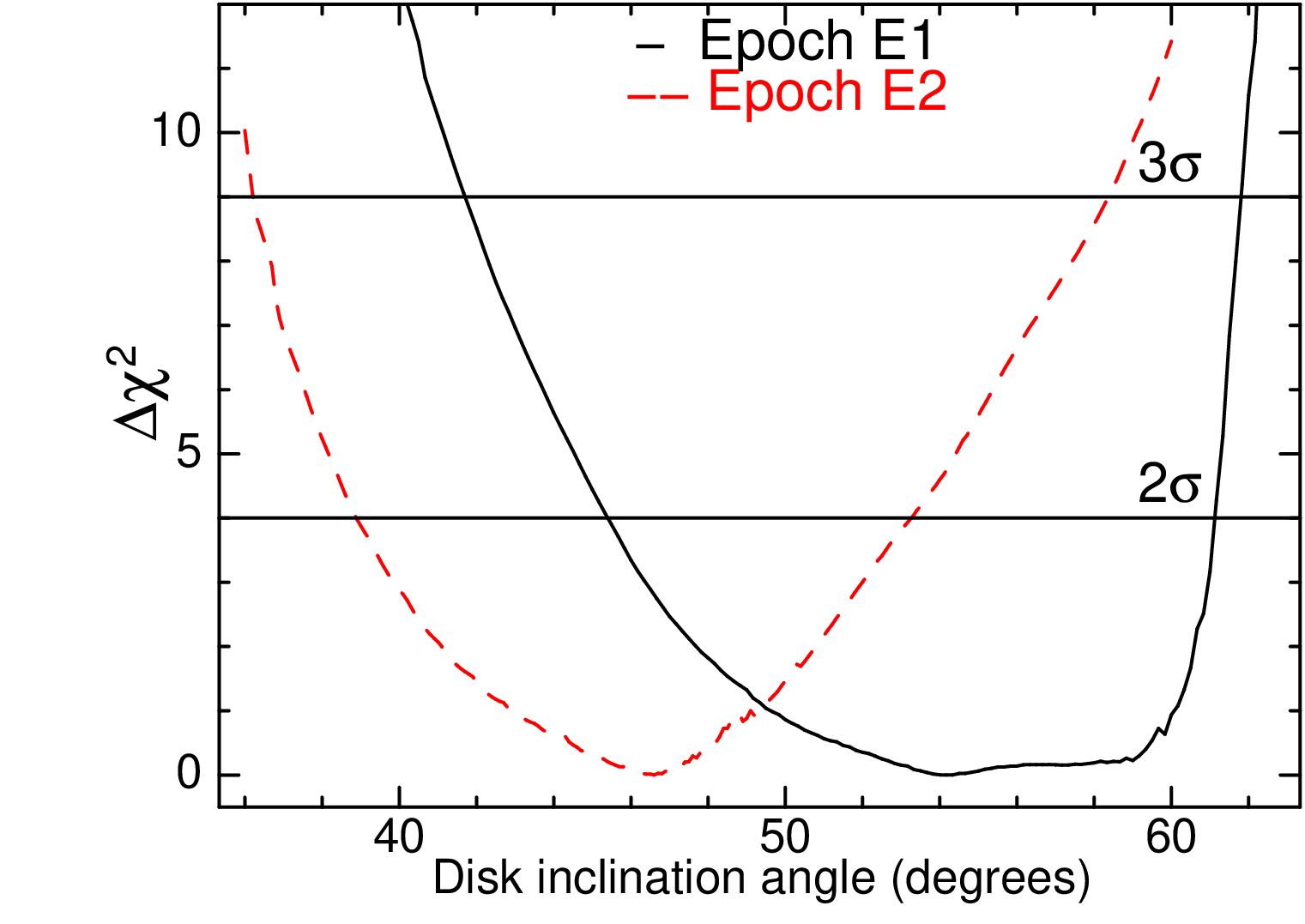}\vspace{2em}
\caption{Variations of $\Delta\chi^{2}$ as a function of the inner accretion disc radius ( R$_{in}$: in units of R$_{ISCO}$, determined from the {\tt relxilllpCp} model) and the disk inclination angle (i). In the {\bf Left panel}, the variation of $\Delta\chi^{2}$ with inner disc radius is shown as observed in Epoch E1 (black solid line) and E2 (red dotted line). It is found that from epoch E1 to E2 within 3$\sigma$ uncertainty, the inner disk radius decreased from 11.92$^{+8.62}_{-11.92}$ (R$_{ISCO}$) to 10.39$^{+9.51}_{-3.02}$ (R$_{ISCO}$). The {\bf Right panel} shows the variations of $\Delta\chi^{2}$ as a function of disk inclination angle. The inclination angle is found to be 54.19$^{+7.43}_{-12.48}$ $^{\circ}$ and 46.59$^{+11.78}_{-10.38}$ $^{\circ}$ for E1 and E2, respectively, within 3$\sigma$ significance. a$^{*}$ is fixed at 0.998.}
\label{fig:fig7}
\end{figure*}

In order to constrain the inner radius of the accretion disc from our best-fit model {\tt (TBabs$\times$(diskbb+relxilllpCp))}, we determined $\Delta\chi^{2}$ using {\tt steppar} command in {\tt XSPEC}. The variation of the resulting $\Delta\chi^{2}$, while changing the inner disc radius as the free parameter between 1 R$_{ISCO}$ and 23 R$_{ISCO}$ for epochs E1 and E2, is illustrated in the left panel of Fig. \ref{fig:fig7}. The 2$\sigma$ and 3$\sigma$ significance levels are shown by the horizontal lines. Within 3$\sigma$ bounds, the value of the inner disc radius for the epochs E1 and E2 are found to be 11.92$^{+8.62}_{-11.92}$ R$_{ISCO}$ and 10.39$^{+9.51}_{-3.02}$ R$_{ISCO}$ or 14.78$^{+10.69}_{-14.78}$ r$_{g}$ and 12.88$^{+8.50}_{-3.74}$ r$_{g}$, respectively. The first epoch (E1) represents only an upper bound. Similarly, we have also constrained the disk inclination angle as shown in the right panel of Fig. \ref{fig:fig7}. The value of the inclination angle is found to be 54.19$^{+7.43}_{-12.48}$ $^{\circ}$ and 46.59$^{+11.78}_{-10.38}$ $^{\circ}$ for E1 and E2, respectively, within 3$\sigma$ bounds.

The un-absorbed total fluxes corresponding to the different continuum (disk and Comptonized emission) components have significantly changed from E1 to E2. For example, the total flux has increased by $\sim5$ \% after considering the proper error propagation. As mentioned earlier, the source was in a relatively softer state at E2 in comparison to E1. The detailed parameters obtained from modeling the broad-band X-ray spectrum (for both epochs E1 \& E2) are summarized in Table \ref{tab:tab2}. We have witnessed significant changes in $\Gamma$, making the overall spectra during E2 relatively softer than during E1. This implies the possibility of successive changes in $\Gamma$, i.e. its evolution during the spectral state change of GRS 1716--249 from E1 to E2. 

Aiming to study this behavior, the overall {\it NuSTAR} light curves during E1 and E2 were subdivided into 8 and 7 fragments, respectively. The same segments were used for investigating the QPOs and the spectral fittings. Due to the unavailability of data at low energies ($<$3.0 keV) during these fragments, we have modeled only the 3.0$-$79.0 keV band. Absorption by neutral hydrogen is modifying the continuum at low energies and in our observations the disk is weak, extending only up to 3.5 keV. Hence, the {\it NuSTAR} data cannot constrain the disk contribution. Therefore, we have kept $n_H$ and T$_{in}$ fixed to the values obtained after fitting the overall average spectrum (listed in Table \ref{tab:tab2}) for both epochs E1 and E2. However, we can assume that certain parameters do not change significantly over a timescale as small as 2 ks. Therefore, we have fixed R$_{in}$ and A$_{Fe}$ to their respective values obtained after fitting the overall average spectrum of epoch E1 and E2, as detailed in Table \ref{tab:tab2}.

\begin{figure*}
\centering
\includegraphics[scale=0.34,angle=-90]{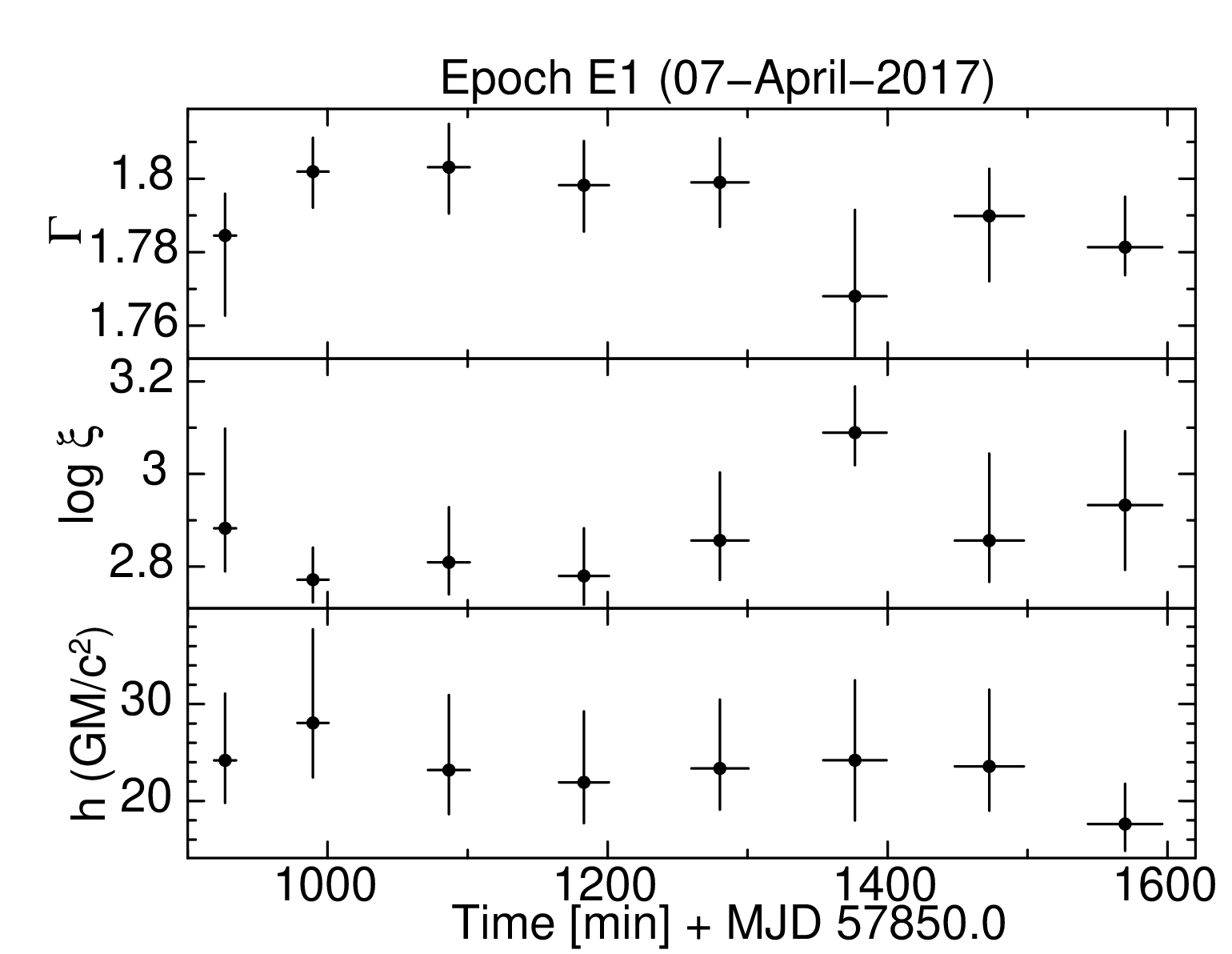}
\includegraphics[scale=0.34,angle=-90]{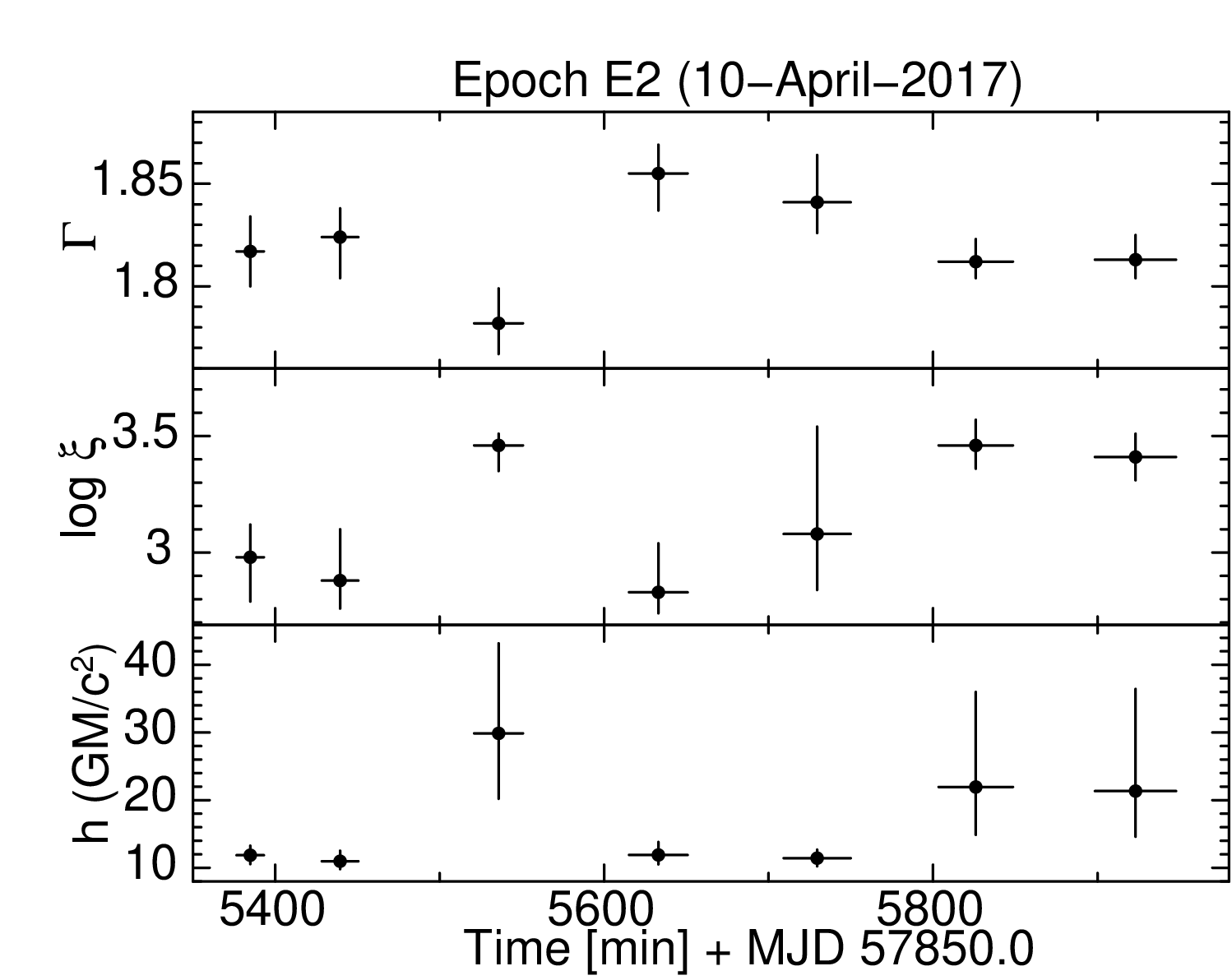}\vspace{1.5em}
\caption{Comparative study of time-resolved energy spectroscopy between E1 and E2. The top panel shows the variation of the photon index ($\Gamma$), the second panel presents the variation of the accretion disk ionization parameter (log $\xi$) and the variation of lamp-post height (h) is shown in the third panel. a$^{*}$ is fixed at 0.998.}
\label{fig:fig8}
\end{figure*}

The results from the above mentioned time-resolved spectral fitting are shown in Fig. \ref{fig:fig8} and are tabulated in Table \ref{tab:tab3}. Note that the error bars shown in Fig. \ref{fig:fig8} represent 2$\sigma$ errors. This clearly shows that the spectral parameters, namely $\Gamma$ and log $\xi$ derived for both E1 and E2 show a hint of variation within 2$\sigma$ significance. $\Gamma$ shows an overall softening from epoch E1 to E2. On the other hand, the lamp-post height (h) is almost invariable during both epochs (E1 \& E2) within 2$\sigma$ uncertainties. When we compare the spectral parameters obtained from time-resolved spectrum fitting and average spectrum fitting, we see that there is some variation in the values of $\Gamma$ and log $\xi$. Some $\Gamma$ values are above the average values, whereas the log $\xi$ values are below the average value. This variation in $\Gamma$ and log $\xi$ could be due to the difference in the energy range between the average spectrum and the time-resolved spectrum. For the average spectrum, we performed a fit in the energy range 0.5--79.0 keV, whereas we carried out time-resolved spectrum fitting in the energy range 3.0--79.0 keV, as we do not have Swift/XRT data for each time interval. A systematically lower log $\xi$ corresponds to a larger $\Gamma$ to produce a similar fit. These two parameters are anti-correlated to some degree and hence, if $\Gamma$ systematically increases, the ionization parameter will decrease \citep{Garcia2013, Choudhury2017}.

\subsection{Correlation Study}

The correlations between the model parameters derived from the time-resolved spectroscopy and the QPO frequency ($\nu_{QPO}$) were also studied. The Pearson correlation test is performed to quantify the correlation, using the following definition:
\begin{itemize} 
\item Pearson Correlation coefficient \citep{Pearson1920}
\begin{equation}
r = \frac{\Sigma(x-\bar{x})(y-\bar{y})}{\sqrt{\Sigma(x-\bar{x})^2}\sqrt{\Sigma(y-\bar{y})^2}}
\end{equation}
where $\bar{x}$ and $\bar{y}$ are the mean of the two series x \& y. 
\end{itemize}
The p-value (significance level) of the correlation is determined by the T-test given by
      \begin{equation}
      t = \frac{r}{\sqrt{1-r^2}}\sqrt{n-2}
      \end{equation}
As interpreted from the T-distribution table, the p-values (p) $\le$ 0.05 indicate a strong correlation.      

The above mentioned test reveals a strong correlation between $\Gamma - \nu_{QPO}$, as displayed in Fig. \ref{fig:fig9}. The corresponding Pearson's coefficient is also shown in the plot. A strong positive correlation is observed for $\Gamma - \nu_{QPO}$ with r = 0.70 (p = 8.78 $\times 10^{-4}$).
 
\begin{table*}
\centering
 \caption{Spectral parameters obtained through time-resolved analysis of {\it NuSTAR} data}
\begin{center}
\scalebox{0.85}{
\begin{tabular}{ |l|l|c|c|c|c|c|c|c|c|c|c|c| }
\hline
\hline
   &   & &  & &  & Epoch E1 &  &   & \\ 
\hline
\hline
Time$^\dagger$ & Exp. & $\nu_{\mathrm{knee}}$ & $\nu_{QPO}$ & Q-factor & $\sigma_{QPO}$ & $\chi^{2}$/dof & h & $\Gamma$ & log $\xi$ & $\chi^{2}$/dof & Total  \\
(min) & (ks) & (Hz) & (Hz) & & & (timimg) & (GM/c$^{2}$) &  & (\ log[\ erg cm s$^{-1}$]\ )\ & (spectral) & Flux$^{\dagger\dagger}$\\ 
\hline
\hline
926.76 & 2.1 & 0.13$^{+0.02}_{-0.02}$ & 1.13$^{+0.01}_{-0.01}$ &  11.57$^{+2.84}_{-2.73}$ & 6.9 & 49.11/42 & 24.19$^{+6.89}_{-4.39}$ & 1.785$^{+0.011}_{-0.022}$ & 2.88$^{+0.22}_{-0.09}$ & 813.44/722 & 12.95$^{+0.04}_{-0.05}$\\
989.54 & 2.2 & 0.15$^{+0.02}_{-0.02}$ & 1.18$^{+0.01}_{-0.01}$ & 8.28$^{+1.91}_{-1.67}$ & 6.4 & 52.88/42 & 28.06$^{+9.68}_{-5.63}$ & 1.802$^{+0.009}_{-0.010}$ & 2.77$^{+0.07}_{-0.05}$ & 764.64/722 & 12.81$^{+0.04}_{-0.04}$\\
1086.59 & 2.2 & 0.15$^{+0.02}_{-0.03}$ & 1.18$^{+0.02}_{-0.01}$ &  8.02$^{+1.81}_{-1.66}$ & 6.3 & 47.88/42 & 23.18$^{+7.76}_{-4.55}$ & 1.803$^{+0.012}_{-0.013}$ & 2.81$^{+0.12}_{-0.07}$ & 675.19/722 & 12.79$^{+0.04}_{-0.04}$\\
1183.08 & 2.2 & 0.12$^{+0.02}_{-0.02}$ & 1.19$^{+0.02}_{-0.02}$ & 5.71$^{+1.41}_{-1.16}$ & 4.4 & 53.81/42 & 21.91$^{+7.34}_{-4.21}$ & 1.798$^{+0.012}_{-0.013}$ & 2.78$^{+0.10}_{-0.06}$ & 738.31/722 & 12.80$^{+0.04}_{-0.04}$\\
1280.21 & 2.2 & 0.14$^{+0.02}_{-0.02}$ & 1.23$^{+0.02}_{-0.02}$ & 9.22$^{+3.21}_{-3.24}$ & 4.5 & 49.11/42 & 23.36$^{+7.11}_{-4.26}$ & 1.799$^{+0.012}_{-0.012}$ & 2.86$^{+0.15}_{-0.08}$ & 791.99/722 & 12.71$^{+0.04}_{-0.04}$\\
1376.69 & 2.3 &  0.15$^{+0.02}_{-0.02}$ & 1.17$^{+0.01}_{-0.01}$ & 7.78$^{+1.71}_{-1.41}$ & 6.8 & 39.56/42 & 24.19$^{+8.26}_{-6.21}$ & 1.768$^{+0.024}_{-0.016}$ & 3.09$^{+0.10}_{-0.07}$ & 757.02/722 & 12.83$^{+0.04}_{-0.02}$\\
1472.66 & 2.2 & 0.14$^{+0.02}_{-0.02}$ & 1.11$^{+0.02}_{-0.02}$ & 5.85$^{+1.42}_{-1.15}$ & 5.9 & 52.01/42 & 23.57$^{+7.93}_{-4.58}$ & 1.790$^{+0.013}_{-0.012}$ & 2.86$^{+0.19}_{-0.09}$ & 720.65/722 & 12.88$^{+0.04}_{-0.04}$\\
1569.67 & 2.2 & 0.16$^{+0.02}_{-0.02}$ & 1.16$^{+0.01}_{-0.01}$ & 9.12$^{+2.10}_{-1.77}$ & 6.6 & 51.95/42 & 17.62$^{+4.15}_{-2.77}$ & 1.781$^{+0.014}_{-0.008}$ & 2.93$^{+0.16}_{-0.14}$ & 751.44/722 & 12.84$^{+0.04}_{-0.04}$\\
\hline
\hline
   &  &  &  &  &  & Epoch E2 &  &  &   \\
\hline
\hline
5384.85 & 2.2 & 0.15$^{+0.02}_{-0.02}$ & 1.45$^{+0.02}_{-0.03}$ & 8.52$^{+1.99}_{-2.32}$ & 5.5 & 48.97/44 & 11.86$^{+1.40}_{-1.30}$ & 1.817$^{+0.017}_{-0.017}$ & 2.98$^{+0.14}_{-0.19}$ & 729.60/722 & 12.03$^{+0.04}_{-0.04}$\\
5439.51 & 2.3 & 0.11$^{+0.02}_{-0.03}$ & 1.45$^{+0.01}_{-0.01}$ &  10.90$^{+2.14}_{-1.81}$ & 8.1 & 56.15/44 & 10.97$^{+1.55}_{-1.17}$ & 1.824$^{+0.014}_{-0.020}$ & 2.88$^{+0.22}_{-0.12}$ & 819.55/722 & 11.95$^{+0.04}_{-0.04}$\\
5535.86 & 2.3 & 0.20$^{+0.02}_{-0.02}$ & 1.47$^{+0.02}_{-0.02}$ & 7.68$^{+1.57}_{-1.48}$ & 7.7 & 41.35/44 & 29.87$^{+13.31}_{-9.66}$ & 1.782$^{+0.017}_{-0.015}$ & 3.46$^{+0.05}_{-0.11}$ & 744.96/722 & 11.88$^{+0.04}_{-0.04}$\\
5632.99 & 2.2 & 0.18$^{+0.02}_{-0.02}$ & 1.53$^{+0.01}_{-0.02}$ & 10.14$^{+2.48}_{-2.25}$  & 6.4 & 50.05/44 & 11.91$^{+1.91}_{-1.39}$ & 1.855$^{+0.014}_{-0.018}$ & 2.83$^{+0.21}_{-0.09}$ & 823.90/722 & 11.79$^{+0.04}_{-0.04}$\\
5729.58 & 2.3 & 0.16$^{+0.02}_{-0.02}$ & 1.65$^{+0.02}_{-0.03}$ & 7.38$^{+2.11}_{-1.63}$ & 5.5 & 45.20/44 & 11.42$^{+1.26}_{-1.19}$ & 1.841$^{+0.023}_{-0.015}$ & 3.08$^{+0.46}_{-0.24}$ & 700.51/722 & 11.60$^{+0.04}_{-0.04}$\\
5825.97 & 2.2 & 0.21$^{+0.02}_{-0.02}$ & 1.67$^{+0.03}_{-0.02}$ & 9.38$^{+3.58}_{-3.40}$ & 4.4 & 50.62/44 & 21.95$^{+14.05}_{-7.08}$ & 1.812$^{+0.011}_{-0.008}$ & 3.46$^{+0.11}_{-0.10}$ & 791.98/722 & 11.61$^{+0.04}_{-0.04}$\\
5923.12 & 2.0 & 0.24$^{+0.02}_{-0.02}$ & 1.60$^{+0.05}_{-0.03}$ & 15.05$^{+14.19}_{-15.05}$ & 2.1 & 52.05/44 & 21.35$^{+15.08}_{-6.76}$ & 1.813$^{+0.012}_{-0.009}$ & 3.41$^{+0.10}_{-0.10}$ & 714.37/722 & 11.56$^{+0.04}_{-0.04}$\\
\hline
\hline
\label{tab:tab3}
\end{tabular}}\\
{\bf $\dagger$} : Time since MJD 57850.0;  {\bf $\dagger\dagger$} : Flux in energy band 3.0--79.0 keV shown in unit of ($\times$ 10$^{-9}$) ergs cm$^{-2}$ s$^{-1}$ \\
Exp. : Exposure Time; $\nu_{noise}$ : Peak frequency of broad Lorentzian noise\\
$\nu_{QPO}$ : Peak frequency of QPO; $\sigma_{QPO}$ : Significance of detection\\
$\Gamma$ : Photon Index;  log $\xi$ : Accretion disk ionization parameter\\
h : the lamp-post height;  a$^{*}$ is fixed at 0.998
\end{center}
\end{table*}

\begin{table*}
\centering
 \caption{Model parameters from simultaneous {\it Swift}/XRT (0.5--8.0 keV) and {\it NuSTAR} (3.0--79.0 keV) spectral fitting. The model which provides the best fit is {\tt TBabs$\times$(diskbb+relxill(lp)Cp)}. 2$\sigma$ errors are quoted. This model resulted in $\chi^2$/dof= 949.04/906 and 1097.55/906 for E1 and E2, respectively. Fig. \ref{fig:fig6} shows the fitted energy spectrum along with the model constituents and residuals for both observations.}
\begin{center}
\scalebox{1.0}{%
\begin{tabular}{ |l|l|l|l| }
\hline
\hline
Component & Parameter & Epoch & Epoch  \\
 &  & E1 & E2 \\ 
\hline
\hline
TBABS & N$_{H}$ ($\times 10^{22} cm^{-2}$) & 0.62$^{+0.02}_{-0.02}$ & 0.59$^{+0.01}_{-0.01}$\\
\hline
diskbb & T$_{in}$(keV) & 0.70$^{+0.06}_{-0.04}$ & 0.59$^{+0.02}_{-0.02}$ \\
  & Norm. & 132.18$^{+53.73}_{-49.11}$ & 437.70$^{+105.49}_{-89.48}$ \\
\hline
relxill(lp)Cp & h (GM/c$^{2}$) & 23.77$^{+16.02}_{-10.84}$ & 12.75$^{+7.30}_{-5.11}$ \\
  & a$^{*}$ (cJ/GM$^{2}$) & 0.998$^f$ & 0.998$^f$ \\
  & i (degrees) & 54.19$^{+6.77}_{-7.89}$ & 46.59$^{+3.83}_{-5.85}$ \\
  & R$^{\dagger}_{in}$ (R$_{ISCO}$) & 11.92$^{+4.78}_{-11.92}$ & 10.39$^{+4.08}_{-2.89}$ \\
  & $\Gamma$ & 1.768$^{+0.013}_{-0.006}$ & 1.812$^{+0.006}_{-0.006}$ \\
  & log $\xi$ (\ log[\ erg cm s$^{-1}$]\ )\ & 3.07$^{+0.06}_{-0.03}$ & 3.19$^{+0.28}_{-0.07}$\\
  & A$_{Fe}$ (solar) & 0.79$^{+0.11}_{-0.05}$ & 1.01$^{+0.34}_{-0.08}$ \\
  & Norm. ($\times$ 10$^{-2}$) & 3.11$^{+0.56}_{-0.32}$ & 3.17$^{+0.85}_{-0.34}$ \\
\hline
 & F$_{Total}$ ($\times$ 10$^{-9}$ ergs cm$^{-2}$ s$^{-1}$) & 14.85$_{-0.16}^{+0.16}$ & 15.59$_{-0.14}^{+0.14}$ \\
 & F$_{diskbb}$ ($\times$ 10$^{-9}$ ergs cm$^{-2}$ s$^{-1}$) & 0.49$_{-0.05}^{+0.05}$ & 0.89$_{-0.07}^{+0.08}$ \\
 & F$_{relxill}$ ($\times$ 10$^{-9}$ ergs cm$^{-2}$ s$^{-1}$) & 14.40$_{-0.15}^{+0.15}$ & 14.88$_{-0.14}^{ +0.14}$ \\
\hline
 & $\chi^{2}$/dof & 949.04/906 & 1097.55/906 \\
\hline
\hline
\label{tab:tab2}
\end{tabular}}\\
{\bf $\dagger$} : Inner disk radius; 
{\bf $^f$} tags imply that the specific parameter was frozen to these values while fitting the spectra\\
{\bf $-$} : Note that all the errors are calculated using {\tt error} command in {\tt XSPEC}.\\
{\bf $-$} : The flux values shown in this table are unabsorbed and calculated for the energy band 0.5$-$79.0 keV.
\end{center}
\end{table*}

\section{Discussion and Conclusions \label{sec:discussion}}

In this work, we present a broad-band X-ray study of the black hole candidate X-ray binary GRS 1716--249 during its 2016--2017 outburst in the energy range 0.5$-$79.0 keV, using {\it Swift}/XRT and {\it NuSTAR} FPMA and FPMB data on two different occasions. The joint spectral analysis shows the presence of a broad iron line and reflection hump around 30 keV, which can be well modeled with the state-of-art relativistic reflection model {\tt relxilllpCp}. We constrained the inner disk radius for both epochs and found that the inner disk tends to moves inward with an increase in the mass accretion rate. A low-frequency QPO is observed at $1.20\pm0.04$ Hz during the first epoch E1 and the QPO frequency is shifted to $1.55\pm0.04$ Hz in the second epoch E2. The time-resolved analysis reveals that there is a hint of variation in the QPO frequency during the second epoch (E2). We observed a strong positive correlation between the QPO frequency and the power-law index.   

The current work reports for the first time the {\it NuSTAR} detection of a low-frequency QPO at $\sim 1.20\pm0.04$ Hz in the LMXB GRS 1716--249. Earlier studies from the 1993 outbursts of GRS 1716--249 detected a QPOs at $\sim$0.04 Hz which slowly drifted to 0.3 Hz at the end of the observation \citep{van96}, but there were no QPO detections at frequencies $\ge$1 Hz. 

We observed that the QPO frequency ($\nu_{QPO}$) increases significantly as the source moves from epoch E1 to E2 and the total flux follows the same trend, which shows a positive correlation between the flux and $\nu_{QPO}$ \citep{Ingram2011}. A positive correlation has also been observed between the photon-index and $\nu_{QPO}$. The changing QPO frequency could be connected to the inner edge of the accretion disk \citep{Takizawa1997}. If the total flux is connected to the mass accretion rate, this signifies that the inner edge of the accretion disk may move inward with an increasing mass accretion rate. 

\begin{figure}
\centering 
\includegraphics[scale=0.32,angle=-90]{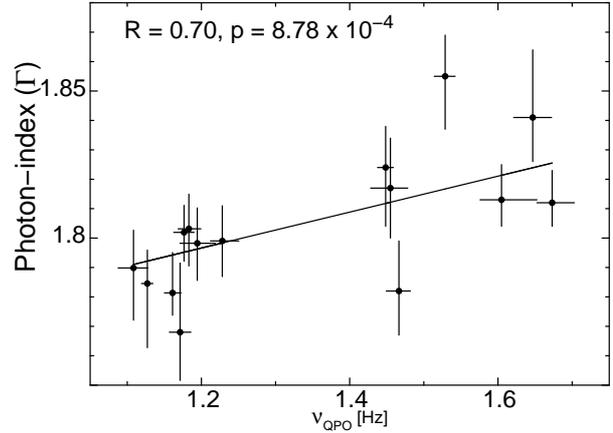}\vspace{1.5em}
\caption{Variation of the photon-index ($\Gamma$) with QPO frequency. The solid line shows the best fit linear dependence. The Pearson ($r$) correlation coefficients for $\nu_{QPO}-\Gamma$ is 0.70 (p= 8.78$\times$10$^{-4}$).}
\label{fig:fig9}
\end{figure}

The correlation between $\Gamma$ and $\nu_{QPO}$ has been beautifully demonstrated by the time-resolved spectroscopy over both epochs (E1 \& E2). The gradual increase of $\Gamma$ over the total 15 segments (Table \ref{tab:tab3}) from E1 and E2, signifies a gradual successive decrease in the non-thermal emission. The cross-correlation study between $\nu_{QPO}$ and the power-law index $\Gamma$ (Fig. \ref{fig:fig9}) clearly establishes a scenario, where the drifting of $\nu_{QPO}$ towards higher frequencies and the increase in $\Gamma$, correspond to a decrease of the non-thermal flux (96.97 \% to 95.45 \%) with an increase of the disk flux (3.30\% to 5.71 \%) as evident from Table \ref{tab:tab2}. Therefore, it is evident that the Comptonizing plasma is diminishing with the disk entering into the soft state.

A strong correlation between $\Gamma$ and $\nu_{QPO}$ has already been observed in a number of X-ray binaries, for example 4U 1608$-$52 and 4U 0614$+$091 \citep{Kaaret1998}, XTE J1550$-$564 and GRO J1655$-$40 \citep{Sobczak2000}, Cyg X-1 \citep{Shaposhnikov2006} and Cyg X-2 \citep{Titarchuk2007}. \citet{Sobczak2000} observed a positive correlation for XTE J1550$-$564 and a negative correlation for GRO J1655$-$40. The authors explained that an increase in the mass accretion rate increases the QPO frequency, and the contribution from the power-law should be more than 20 \% for QPOs to be present. \citet{Sobczak2000} have also suggested that the opposite correlation in XTE J1550$-$564 and GRO J1655$-$40 could be due to different regions of QPO generation in the two sources. To explain the correlations, \citet{Titarchuk2004} proposed the transition layer (TL) model. According to this model, a compact bounded coronal region is formed as a natural consequence of the adjustment of the Keplerian disk flow to the innermost sub-Keplerian boundary conditions near the central region. It ultimately ends up forming a TL between the adjustment radius and the innermost boundary. However, this mechanism is unable to produce the inclination-dependent QPOs obtained by \citet{Motta15}.

The observed correlation can be successfully explained on the basis of the Lense-Thirring (LT) precession model \citep{Ste98, Ing09} and the truncated disc model \citep{Esi97, Pou97, Done2007}. When the mass accretion rate shows rapid growth, the truncated disk radius slowly starts moving towards the compact object, hence leading to an increase of the LT precession frequency. The QPO frequency and its evolution are governed by the size of as well as the fluctuations in the truncation radius (increase in QPO frequency represents a decrease of the truncation radius)\citep{Motta17}. The spectral changes during the source transition can also be explained in this framework. When the mass accretion rate increases, the outer disk gradually starts moving towards the compact object. The disk component becomes stronger and causes greater cooling of the hot inner flow by the cool photons from the disk, resulting in a soft spectrum \citep{Motta15, Zhang2015}.

In our spectral analysis, we can estimate the inner radius of the accretion disk (R$_{in}$) in two different ways. The {\tt relxilllpCp} model directly provides R$_{in}$, which is found to be 14.78$^{+10.69}_{-14.78}$ (r$_{g}$) and 12.88$^{+8.50}_{-3.74}$ (r$_{g}$), for epoch E1 and E2, respectively. R$_{in}$ can be calculated from the normalization of the {\tt diskbb} model. Using the disc inclination obtained from our spectral analysis, taking the mass to be 4.9 M$_\odot$ \citep{Mas96} and the distance to be 2.4 kpc \citep{del94}, R$_{in}$ corresponding to epoch E1 and E2 found to be 0.66$^{+0.13}_{-0.13}$ (r$_{g}$) and 1.12$^{+0.12}_{-0.13}$ (r$_{g}$), respectively. We find that the R$_{in}$ values calculated from the normalization of the {\tt diskbb} model are smaller than the values estimated from the {\tt relxilllpCp} model. One possible reason for the discrepancy in the R$_{in}$ values is that the {\tt relxilllpCp} model is not self-consistent as it uses the {\tt nthComp} model with seed photon temperature (T$_{bb}$) fixed at 0.05 keV. This value of T$_{bb}$ is uncharacteristically low for a stellar mass black hole accretion disk. As a result, a higher temperature component becomes necessary to represent a part of the existing cold disk. Secondly, \citet{Merloni2000} carried out a critical analysis of the usual interpretation of the multicolor disc model parameters for black hole candidates in terms of the inner radius and temperature of the accretion disc. The authors have reported that the {\tt diskbb} model underestimates the inner disk radius. \citet{Merloni2000} suggested that when the disk contribution is very low and the spectrum is mostly dominated by non-thermal photons, the radius inferred by {\tt diskbb} is inaccurate. They have also explained that it is very difficult to determine the exact shape of the accretion disk spectrum for the aforementioned case. \citet{Kubota1998} suggested a correction factor to improve the value of R$_{in}$ calculated from {\tt diskbb}.

The presence of type--C QPO and the broad-band spectral fitting at both epochs (E1 \& E2) confirms that the power-law component dominates over the disk contribution, hence confirming the intermediate state of GRS 1716--249. Here, the soft photons from the accretion disk are Comptonized to higher energies via a hot corona of thermal electrons through the inverse Compton effect \citep{Haardt1991, Haardt1993}. The two epochs have shown significant spectral changes leading to a successive softening of the source as inferred from the broad-band spectrum at E1 and E2. The broad Fe-K$\alpha$ line at 6.4 keV is an indication of a spinning black hole \citep{Miller2009}. The observed iron line is extended from nearly 6 to 8 keV along with a dip around 11 keV. This shows that both wings of the Fe-K$\alpha$ line are stretched. It is believed that the broadening of the red wing might be due to Doppler and General Relativity (GR) effects, while the blue wing is elongated because of scattering of the photons within the hot inner flow \citep{Fabian2000, Miller2007, Miller2017}. 

Reflection features have already been detected in a number of X-ray binaries. \citet{Done1999} have significantly detected the relativistic smearing of the reflected spectrum in Cyg X--1. Using the X-ray spectrum, they have simultaneously constrained the ionization state and the percentage of relativistic smearing. They analyzed observations from three different missions namely: {\it EXOSAT} GSPC, Ginga and {\it ASCA}. They implemented the {\tt PEXRIV} model in {\tt XSPEC} to fit the reflection spectrum, and the {\tt DISCLINE} model in {\tt XSPEC} is used to account for the relativistic smearing. They fitted the spectrum for different values of the inclination angle and iron abundance. During the entire analysis, the photon-index ranged from 1.44 to 1.91. They also reported $\xi$ and R$_{in}$ for various observations. Our results are found to be consistent with their results, and during our observations, $\Gamma$ varies from 1.80--1.93. On many occasions, the values of the inner disk radius and accretion disk ionization parameter as reported by \citet{Done1999}, are found to be close to the values observed in this work. 

In a separate study, \citet{Miller2002} resolved the Fe K$\alpha$ line region in Cyg X-1 through spectral analysis. The {\it Chandra} X-ray observatory detected the source in an intermediate spectral state during these observations. The authors observed a narrow line around 6.42 keV along with a broad line feature at $\sim$ 5.82 keV and a smeared edge at 7.3 keV. These results support our findings, as the study reported a photon index around 1.8 which is close to the value we found during E1. The value of R$_{in}$ reported in that study is similar to the value of R$_{in}$ found during E1 in this work. Two different lines and a smeared edge are not observed in our study, either because of the superior spectral resolution of {\it Chandra} in comparison to {\tt NuSTAR} \citep{Canizares2000, Harrison2013}, or because they may not be present in GRS 1716--249.

Recently, there have been several attempts to find a correlation between QPOs and iron lines to constrain the origin of the QPOs \citep{Miller2005, Ingram2015}. The basic idea behind such an attempt is that both the QPOs and the Fe-line possibly originate from the inner part of the accretion disk. Therefore, they can provide an independent signature of the geometry of the inner disk. It has been found that the variation in the QPO frequency and the QPO phase is correlated with the variation of the iron line centroid frequency \citep{Ingram2016, Ingram2017}. These studies reflect the geometric origin of QPOs. However, our current findings are limited to explore any such correlations between QPO and Fe-line because of the limited extent of our observations. In future work, we plan to use better timing and multi-epoch data from {\it ASTROSAT} \citep{Antia2017, Verdhan2017}, spanning over several months, along with our current results, to study the QPOs with more physical explanations and rigorously updated physical models.

\section*{Acknowledgements}
This research has made use of the software and/or data obtained through the High Energy Astrophysics Science Archive Research Center (HEASARC) online service, provided by the NASA/Goddard Space Flight Center and the {\it SWIFT} data center. We thank {\it NuSTAR} team for making {\it NuSTAR} data public. {\it MAXI} data are obtained from {\it MAXI} team, RIKEN, JAXA. The authors gratefully acknowledge the anonymous referee for constructive comments that improved the paper. The authors are also thankful to PI's for proposing these observations. JC and SC are also grateful to Prof. H. M. Antia, Prof. A. R. Rao and Prof. S. Bhattacharyya for constructive discussions about diagnostics and interpretations.

\bsp
\label{lastpage}

\end{document}